**Main Manuscript for**

Superior Polymeric Gas Separation Membrane Designed by Explainable Graph Machine Learning

Jiaxin Xu[1,+], Agboola Suleiman[2,+], Gang Liu[3,+], Michael Perez[1], Renzheng Zhang[1], Meng Jiang[3,*], Ruilan Guo[2,*], Tengfei Luo[1,2,*]

Affiliations:

[1] Department of Aerospace and Mechanical Engineering, University of Notre Dame, Notre Dame, IN 46556, United States of America

[2] Department of Chemical and Biomolecular Engineering, University of Notre Dame, Notre Dame, IN 46556, United States of America

[3] Department of Computer Science and Engineering, University of Notre Dame, IN 46556, United States of America

[+] Equal contribution

* Corresponding authors: mjiang2@nd.edu; rguo@nd.edu; tluo@nd.edu

**This PDF file includes**:

    Main Text
    Figures 1 to 4
    Tables 1 to 3

**SUMMARY**

Gas separation using polymer membranes promises to dramatically drive down the energy, carbon, and water intensity of traditional thermally driven separation, but developing the membrane materials is challenging. Here, we demonstrate a novel graph machine learning (ML) strategy to guide the experimental discovery of synthesizable polymer membranes with performances simultaneously exceeding the empirical upper bounds in multiple industrially important gas separation tasks. Two predicted candidates are synthesized and experimentally validated to perform beyond the upper bounds for multiple gas pairs ($O_2/N_2$, $H_2/CH_4$, and $H_2/N_2$). Notably, the $O_2/N_2$ separation selectivity is 1.6-6.7× higher than existing polymer membranes. The molecular origin of the high performance is revealed by combining the inherent interpretability of our ML model, experimental characterization, and molecule-level simulation. Our study presents a unique explainable ML-experiment combination to tackle challenging energy material design problems in general, and the discovered polymers are beneficial for industrial gas separation.

**PROGRESS AND POTENTIAL**

Chemical separation, mainly through inefficient energy-intensive thermal processes, consumes 10-15% of global energy, making it imperative to find low-energy alternatives. Materials-driven, membrane-based separations present a promising solution, as they promise to dramatically drive down the energy, carbon, and water intensity of many traditional thermally driven separation processes. The creation of novel membrane materials with tailorable yet predictable structures and properties holds the key to providing low-energy solutions to some of the world's most challenging and vital separations. However, the development cycles of such materials are usually exceptionally long due to the trial-and-error strategy traditionally used. This study leverages a self-explainable graph-augmented and imbalanced machine learning technique to design two exceptional polymers for the efficient separation of multiple industrially critical gas pairs.

**MAIN TEXT**

**INTRODUCTION**

The creation of novel membrane materials with tailored properties holds the key to providing low-energy solutions to many of the separation-related challenges facing humanity in health, energy, and sustainability. Membrane-based separation technologies promise to dramatically drive down the energy-, carbon- and water-intensity of many traditional thermally driven separation processes (1). Among different separation applications, gas separations are central to many technological innovations and advancements in clean energy industries (e.g., $H_2$ purification and air separation) and climate change remediation (e.g., carbon capture). Membranes take advantage of material selectivity rather than thermal energy to perform separations (2–5), with fast and selective transport being their crucial attributes. High selectivity yields high product purity and low operation costs, while high permeability reduces membrane area and capital and energy costs (2, 3, 5). Polymer membranes are of particular interest given their unique advantages, such as lower cost and greater adaptivity. However, polymer membranes are frequently challenged by a tradeoff between permeability and selectivity, which is known as the "upper bound" line in a plot of selectivity versus permeability (6–8) for different gas pairs (e.g., Fig. 1a).

To transcend this "upper bound", novel strategies, such as polymers of intrinsic microporosity (PIM) (*9–12*), thermally rearrangement polymers (*12–16*), and doping (*17–20*), have been explored, but they are still far away from large-scale production. State-of-the-art gas separation polymer membranes are mainly based on polyimides, but most of their properties are below the upper bound, so are those of commercial polymer membranes (Fig. 1a). Despite the advancement of theoretical and modeling tools (*21–26*), developing a new polymer that has high gas separation performance for target gas pairs is still very time-consuming.

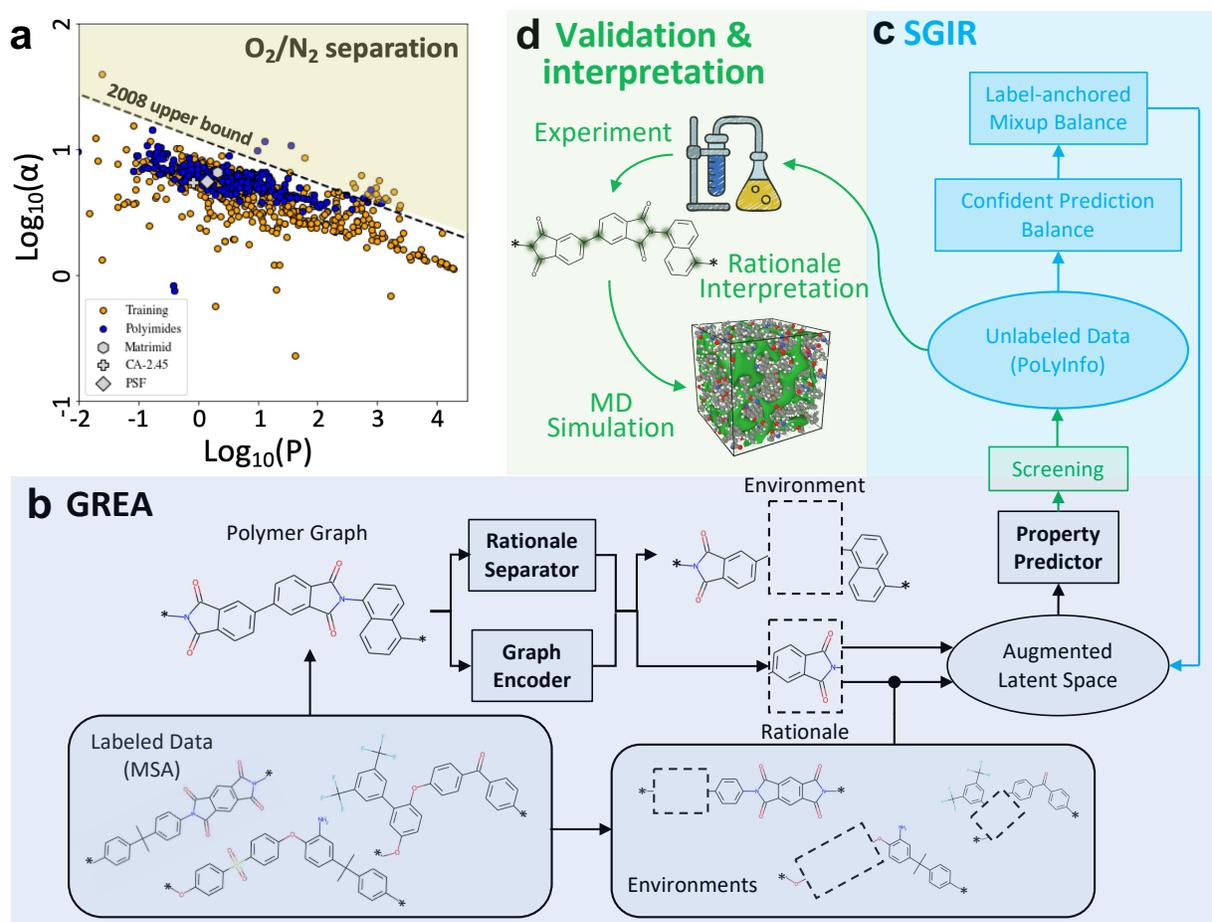

**Figure 1. Gas separation performance upper bound and the GREA-SGIR method for designing polymers above this upper bound.** (**a**) Plot showing the 2008 upper bound correlation for $O_2/N_2$ separation using the labeled training data, where all the polyimides in the training data are colored in blue, the rest of the training data are colored in orange, and three representative commercial polymers in the gas separation industry (Matrimid®, CA-2.45, and PSF) are colored in grey. (**b**) Flow diagram of GREA. The small training data is first represented and augmented by GREA through rationale-environment separation and/or environment-based augmentation. (**c**) The imbalanced training data is further balanced by the SGIR algorithm through two-step balancing, confident prediction balancing and label-anchored mixup balancing utilizing the large unlabeled polymer structure dataset. After the model is trained, screening on the unlabeled dataset is carried out to identify promising candidates for various gas separation tasks. (**d**) Experiment synthesis and testing are then conducted to validate the prediction results. Rationale interpretations from GREA and MD simulations are performed to gain chemical insights from the predicted results.

Data-driven approaches leveraging machine learning (ML) algorithms promise to help scan large chemical spaces efficiently to identify promising candidates. As early as 1994, Wessling et al. (*27*) proposed an IR spectrum-based deep neural network to study the permeability of $CO_2$. More recently, Zhu et al. (*28*) used a hierarchical fingerprint and trained Gaussian Process Regression (GPR) models based on 315 polymers for major industrial gases. However, the training data have very few points above the "upper bound", therefore GPR predictions are hardly above it either. Barnett et al. (*29*) utilized the daylight-like fingerprint and trained GPR models based on the experimental data of six gases with ~400-700 data for each gas. Using the GPR models, they screened a larger database, identified two promising polymers, and successfully synthesized them. However, the fingerprinting method used to represent polymers made the prediction result unexplainable, and the identified two polymers were only targeted for one separation task ($CO_2$/$CH_4$). Yang et al. (*30*) combined molecular dynamics (MD) simulations and supervised ML to identify promising polymer candidates and were able to explain their prediction using SHAP (Shapley additive explanations) (*31*). While many polymers were predicted to have above-the-upper-bound performance, no experimental validation was carried out. Therefore, using ML to design polymers with performance above the gas separation upper bounds remains a daunting challenge due to the overall limited data availability (small data), extreme data scarcity above the upper bound (imbalanced data), difficulty in explaining the results (model interpretability), and the need of experimental validation.

In this work, we first employ a unique explainable data-augmentation graph neural networks (GNNs), the graph rationalization enhanced by the environment-based augmentations (GREA, Fig.

1b), to tackle the small data problem. The polymer graph is first divided into two parts, the rationale subgraph and the environment subgraph. The GREA method can highlight the essential features in the polymer graph representation through the rationale subgraph so that small data can effectively train the model while minimizing overfitting and improving generalizability. Since the labeled data are limited (500-800 per gas), a semi-supervised framework is used to leverage the much larger unlabeled dataset PoLyInfo (*32*). Semi-supervised learning has been widely demonstrated to enhance model performance in the presence of balanced labeled data (*33*, *34*). However, in cases of imbalanced data, applying semi-supervised learning without considering label imbalance may lead to decreased model performance (*35*), particularly for minority labels located above the upper bound in our study (e.g., Fig. 1a). To address this issue, we leverage a framework known as semi-supervised graph imbalanced regression (SGIR, Fig. 1c), which have proven effective in reducing prediction errors within the under-represented label range through a two-step balancing (*36*). Using the established models, we screen the largest polymer database, PoLyInfo (*32*), and identify two polymers that are predicted to simultaneously out-perform the 2008 upper bounds for three different gas pairs ($O_2/N_2$, $H_2/CH_4$, $H_2/N_2$), including the challenging $O_2/N_2$ separation due to the similarity in their molecular sizes. Both polymers are synthesized and tested experimentally (Fig. 1d). It is worth highlighting that the exceptional $O_2/N_2$ selectivity of one of the polymers, which was initially predicted through our model, has been experimentally confirmed through synthesis and measurement. The resulting selectivity value of 17.48 is particularly noteworthy, which is 1.6-6.7× higher than many existing state-of-the-art polymer membranes (*37*). The ML predictions are found to be 80% accurate in terms of whether their performance is above the upper bound. Meanwhile, experimental characterization and MD simulations are performed to better understand the origin of the high performance of the two polymers. The learned rationale is used to interpret the ML model decision-making process, following which the backbone rigidities are analyzed using the semi-empirical atomistic model. Our study presents a unique self-explainable ML technique to tackle small and imbalanced data problems, which are frequently encountered in materials development missions, and the resulting polymers may be used for high-performance membranes to separate several industrially critical gas pairs.

**RESULTS**

**Machine Learning Models**

The GREA model is implemented to address the data insufficiency issue in the training dataset as well as to provide model interpretability. It is composed of three parts, including a rationale separator, a graph encoder, and a property predictor (Fig. 1b). The rationale separator takes a polymer graph as input (see Methods section for details of polymer representations) and then outputs a probability value of each atom in the graph. These probabilities represent the likelihood of each atom being classified as the rationale of the polymer. A rationale is the causal subgraph structure extracted for accurate and interpretable graph property predictions, in other words, the key dictating the properties. In contrast, the remaining subgraph is referred to as the environment subgraph corresponding to the non-causal part, and the model treats it as "noise" when predicting properties. The environment subgraph is necessary because it can not only provide complementary information to the rationale subgraph, making the entire polymer graph structure chemically valid and complete, but also serve as small perturbations to the identified rationales enabling data augmentation. The training data augmentation feature in GREA is realized by removing the environment subgraph from the rationale subgraph and/or combining it with different polymers' environment subgraphs in the latent space. Here, a two-layer graph isomorphism network (GIN) (*38*) is implemented as the rationale separator. The graph encoder also takes a polymer graph as input, and it outputs the graph representation, which is further projected to a graph label by the property predictor. A five-layer GIN is implemented as the graph encoder and a three-layer multilayer perceptron (MLP) is implemented as the property predictor. Training details are included in the Methods section. The interested reader is referred to Ref. (*39*) and references cited therein for more detailed descriptions. Two other popular GNNs used in polymer property regression, including graph convolutional network (GCN) (*40*) and GIN, are also trained as baselines to the GREA model. Please see supplemental information (SI), Section 1 for details of these models. All models are trained using a semi-supervised framework for graph imbalanced regression (SGIR, Fig. 1c, see Methods for details). For simplicity, we only call these methods by the GNNs part (namely, GREA, GCN, and GIN) in the following text, omitting the SGIR part in the name.

The labeled data are mainly collected from the Membrane Society of Australasia (MSA) (*41*) Polymer Gas Separation Membrane Database, and a thorough data cleaning process is performed. All homopolymers with no gas transport property data recorded from the PoLyInfo database (12,769 data points) are used as unlabeled data. See SI, Section 2 for the description of the two datasets and data cleaning details. Five important industrial gas pairs are studied ($H_2/N_2$, $H_2/CH_4$, $O_2/N_2$, $CO_2/N_2$, and $CO_2/CH_4$). Given the small amount of available data, all labeled data are used for training. As shown in Fig. 2, the dominant majority of labeled data (orange points) lies below the 2008 upper bounds (ranging from 93.0 % to 98.6 % across the five different gas pairs), which would impair the generalization of the trained model to the above-the-upper-bound regime. Randomly splitting the labeled data for the validation set may not include data points above the upper bounds, potentially leading to overfitting to the below-the-upper-bound regime. To enhance the model's applicability in predicting candidates that can potentially have above-the-upper-bound performance, we use data above the upper bound as the validation set instead of the more conventionally used random split of the whole dataset. We make this deliberate selection of validation set to further mitigate the model bias caused by the majority of labeled data and better meet our design goal to discover polymer candidates above these upper bounds. All methods run 10 times independently and the means and standard deviations of the mean absolute error (MAE) on the validation sets are reported in Table 1. For all gases, except $CO_2$, GREA outperforms the other two graph models. For $CO_2$, GREA is the close second next to GCN. The realization of rationale separation and environment replacement augmentation is the key to the success of GREA. It allows the model to focus only on the rationale part of the input polymer graph and be robust to the noise to maximize the performance of the downstream task, which should be especially helpful when training with small data.

**Table 1. Model performance comparison for the prediction of five gas permeability tasks.** Average MAE values with standard deviations on the validation dataset are shown. The best-performing model for each task is bolded. All MAE values are in units of $\log_{10}$Barrer, where 1 Barrer = $1\times10^{-10}$cm$^3$ [STP] cm$^2$/(cm$^3$ s cmHg).

| Gas  | $N_2$ | $O_2$ | $H_2$ | $CH_4$ | $CO_2$ |
|------|-------|-------|-------|--------|--------|
| GREA | **0.571±0.083** | **0.482±0.090** | **0.432±0.113** | **0.588±0.108** | 0.487±0.164 |
| GCN  | 0.650±0.217 | 0.536±0.149 | 0.447±0.064 | 0.609±0.125 | **0.461±0.091** |
| GIN  | 0.797±0.369 | 0.568±0.081 | 0.495±0.088 | 0.596±0.109 | 0.568±0.310 |

**Screening the PoLyInfo Database for Promising Candidates**

The trained GREA models for different target gases are then applied to 12,769 previously synthesized polymers from the PoLyInfo database to predict their gas transport permeabilities. The prediction on this large repository represents a large amount of new polymer gas transport data, which have never been experimentally tested before, providing useful guidance on the design of high-performance polymer membranes. Figure 2 shows the result of gas permeability-selectivity predicted on polymers in the PoLyInfo database (shown in blue circles) for the five gas separation tasks, plotted in the format of the Robeson plot, where the x-axis $\log_{10}P$ is the predicted permeability in the unit of $\log_{10}$Barrer, and the y-axis is the calculated selectivity $\alpha$ derived from the ratio of permeabilities of the gas pairs on the $\log_{10}$ scale. For the screened polymers, the medians from the ensemble of ten independent models are reported. Most of the predicted permeability-selectivity values still lie below the upper bounds for all the studied gas pairs, which is consistent with the training data. However, the predicted areas (blue) also extend beyond the training areas (orange), suggesting that GREA can have certain extrapolatability.

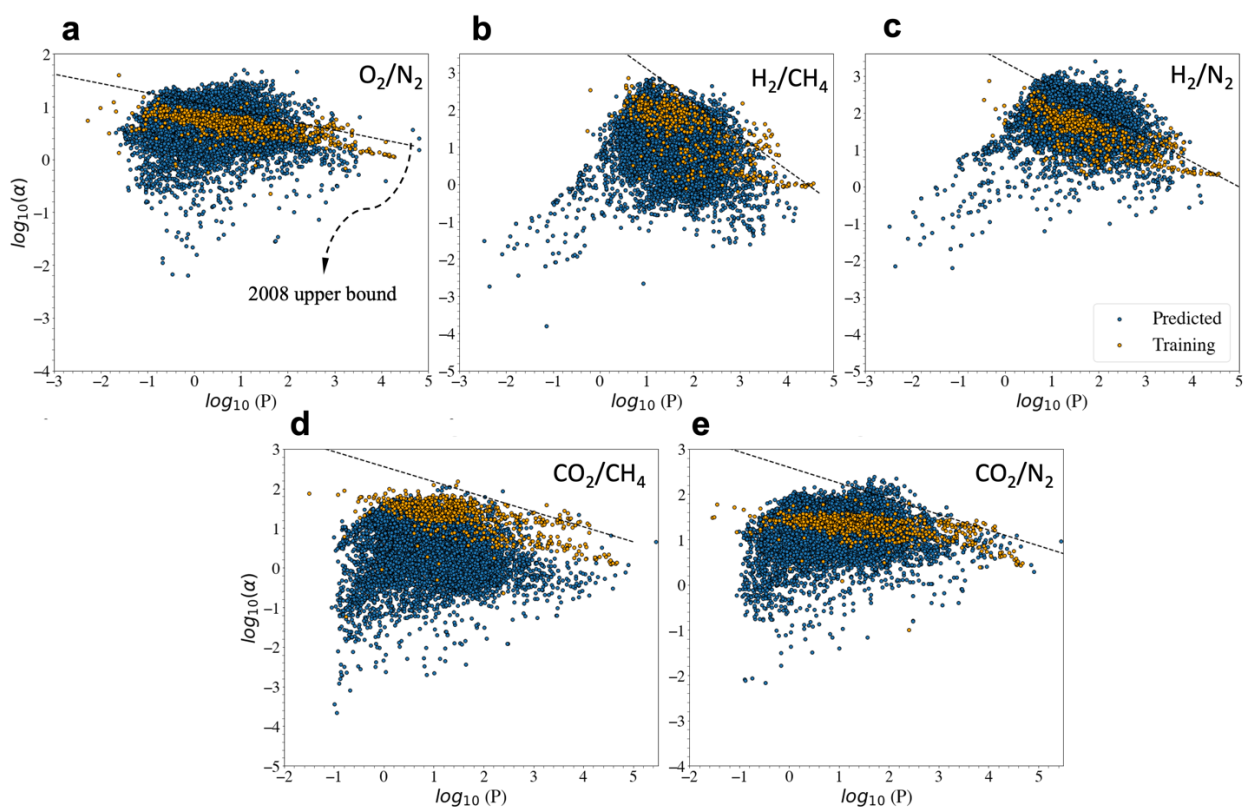

**Figure 2**. **Training and screening for polymer membrane candidates.** Robeson plot visualization for five different separation tasks: **(a)** $O_2/N_2$, **(b)** $H_2/CH_4$, **(c)** $H_2/N_2$, **(d)** $CO_2/CH_4$, and **(e)** $CO_2/N_2$, including the training data (orange circles) and the median values of prediction results on PoLyInfo dataset (blue circles). The 2008 upper bounds for different tasks are shown as black dashed lines. The x-axis $\log_{10}P$ is the predicted permeability in the unit of $\log_{10}$Barrer, and the y-axis is the predicted selectivity $\alpha$ on the $\log_{10}$ scale.

Similar to previously published data for polymer screening (*29*, *30*), a reverse selectivity problem in the prediction results of our ML models is observed. For example, some polymers in the PoLyInfo database are predicted to be $N_2/O_2$ and $CH_4/H_2$ selective, which is not physical. We attribute this problem mainly to the propagation of uncertainty, where the prediction uncertainty on the permeability values is further propagated to the calculated selectivity values. As a result, some of the polymers in our dataset exhibit reverse selectivity. Another reason for the observed reverse selectivity in prediction models is the presence of reverse selective training data. Our analysis of the training data revealed that for $H_2/CH_4$ separation, there were 18 reverse selective training data, mainly comprising glassy polymers such as substituted polyacetylenes (*42–44*), which have very high free volume and gas permeability values. This makes it possible for smaller gas molecules to travel slower, leading to reverse selective behavior. For $O_2/N_2$ separation, six reverse selective training data were identified, and the reverse phenomenon mostly resulted from the data in the MSA database itself. However, since the percentage of such data is relatively small (6/785), we do not expect any significant issues to arise from these training data. This also applies to the cases of $CO_2/CH_4$ (4/672) and $CO_2/N_2$ (1/730). And no reverse selective training data was present for $H_2/N_2$ separation. Moreover, as for $CO_2/CH_4$ and $CO_2/N_2$, the distinctions in condensability and reactivity between $CO_2$ and other light gases further add complexity to the situation and could exacerbate the issue of reverse selectivity prediction.

**Experimental Synthesis and Validation**

Based on the screening results, we select two high-performance candidates for experimental validation. They are poly[(naphthalene-1,5-diamine)-alt-(biphenyl-3,3':4,4'-tetracarboxylic dianhydride)] (PoLyInfo ID: P130093) and poly[isophoronediamine-alt-(biphenyl-3,3',4,4'-tetracarboxylic anhydride)] (PoLyInfo ID: P432352). Their molecular structures are shown in Fig. 3a. The two candidates were chosen since their performances are predicted to be simultaneously above the upper bounds for multiple gas pairs, especially showing exceptional $O_2/N_2$ selectivity. Equally importantly, they are aromatic polyimides, normally the polycondensation product when aromatic dianhydrides are reacted with diamine monomers. Both polyimides exhibit conventional structural constituents, favorable processability, and economic feasibility. This provides us with adequate confidence in their synthesizability. The two polymers are synthesized via conventional polycondensation reactions (see Method section for synthesis details). Their permeabilities for the five different gases are measured and the selectivities for the five gas pairs are calculated (see Method section for gas permeation tests details).

**Table 2. Experimental pure-gas permeation data for the two selected candidates and three commercial polymer membranes.** The permeability coefficients of the two selected candidates were measured using a constant-volume variable-pressure method (*45*) at 35 °C, and the permeabilities of the three commercial polymers were from Ref. (*2*). The selectivity coefficients were calculated based on Eq. (2).

| | *Permeability coefficient (barrer)* | | | | | *Selectivity coefficient* | | | | |
|---|---|---|---|---|---|---|---|---|---|---|
| *Polymer* | $N_2$ | $O_2$ | $H_2$ | $CH_4$ | $CO_2$ | $O_2/N_2$ | $H_2/CH_4$ | $H_2/N_2$ | $CO_2/CH_4$ | $CO_2/N_2$ |
| *P130093* | 0.025 | 0.437 | 23.1 | 0.017 | 1.548 | 17.48 | 1358.82 | 924.00 | 91.06 | 61.92 |
| *P432352* | 0.165 | 1.199 | 54.5 | 0.113 | 5.25 | 7.27 | 482.30 | 330.30 | 46.46 | 31.82 |
| *Matrimid®* | 0.32 | 2.1 | 18 | 0.28 | 10 | 6.56 | 64.29 | 56.25 | 35.71 | 31.25 |
| *CA-2.45* | 0.15 | 0.82 | 12 | 0.15 | 4.8 | 5.47 | 80.00 | 80.00 | 32.00 | 32.00 |
| *PSF* | 0.25 | 1.4 | 14 | 0.25 | 5.6 | 5.60 | 56.00 | 56.00 | 22.40 | 22.40 |

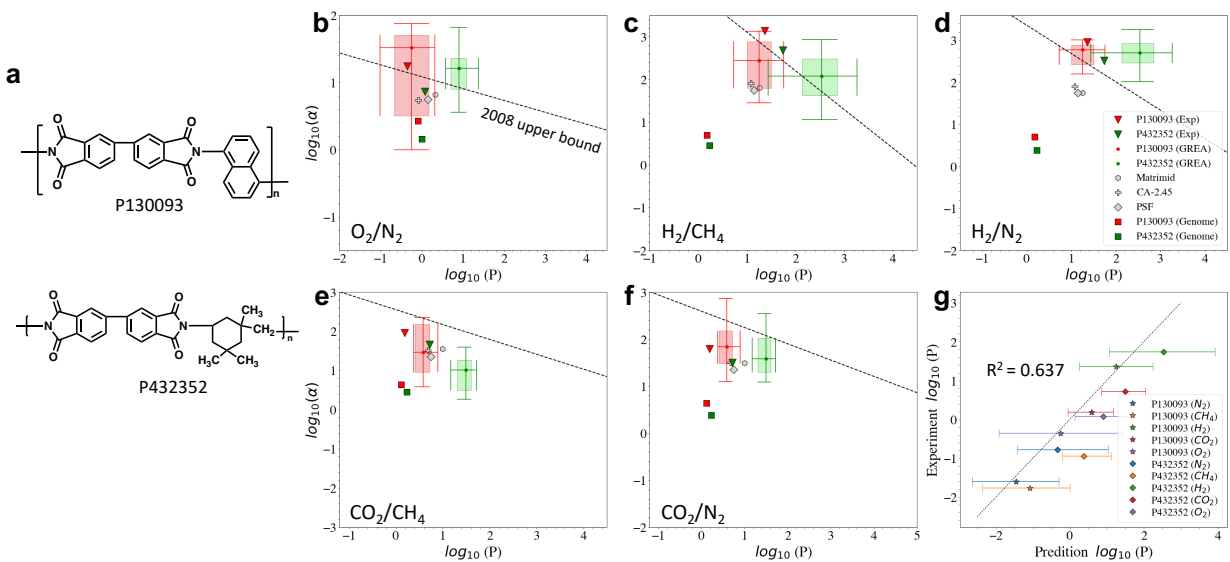

**Figure 3**. **Experimental validation of the two selected polymer candidates. (a)** Repeat units of the selected candidates: P130093 (upper) and P432352 (lower). **(b-f)** Robeson plot visualization for five different separation tasks using data from different sources, including (1) 2D box plots of the 10 independent GREA prediction results of P130093 (red) and P432352 (green) polymers (shaded rectangle box denotes the area within the first and third quantile, the dot in the box denotes the median, and the whiskers extend from the rectangle box by 1.5× the inter-quartile range), (2) the experimental measurements of the P130093 (red) and P432352 (green) polymers (inverted triangle symbols), (3) the prediction results of the P130093 (red) and P432352 (green) polymers from Polymer Genome (square symbols), and (4) experimental results of representative commercial polymer membranes, Matrimid®, CA-2.45, and PSF (light grey symbols). The 2008 upper bounds for different tasks are shown as black dashed lines. The x-axis $log_{10}P$ is the permeability in units of $log_{10}$Barrer, and the y-axis is the calculated selectivity $\alpha$ on the $log_{10}$ scale. **(g)** Predicted permeability using GREA versus experimentally measured permeability of P130093 (star) and P432352 (diamond) for $N_2$ (blue), $CH_4$ (orange), $H_2$ (green), $CO_2$ (red), and $O_2$ (purple). Each circle is the median of 10 independent prediction results, and the error bars are equal to the second or third quantile added with 1.5× the inter-quartile range. The black dashed line represents parity.

The experimental results (Table 2) are in reasonable agreement with the GREA model prediction (Fig. 3b-f). Since predicting above-the-upper-bound candidates is deemed an extrapolative task and the labeled training data size is relatively small, it is not surprising that the prediction uncertainty is relatively high. However, a widely used performance gauge assessing the potential of a polymer material for gas separations is the position of the permeability/selectivity data relative to the upper bounds. Therefore, we consider the accuracy of the models as if they correctly predict the placement of the properties in the permeability-selectivity plot with respect to the upper bounds (i.e., a binary classification with labels of "below" or "above"). According to Fig. 3b-f, the GREA model achieves an accuracy of 8/10, where there are 10 data points in total (two polymers and five separation tasks), and 8 of them are predicted to be correctly located with respect to the upper bounds compared with the experimental results. From only the permeability point of view, Fig. 3g shows the parity plot between the predicted permeabilities and the experimental measurements, revealing a good correlation with $R^2=0.637$. The predicted permeability order for the two candidates is $H_2>CO_2>O_2>N_2>CH_4$, consistent with the order of kinetic diameters of these five gases ($H_2$=2.89 Å, $CO_2$=3.3 Å, $O_2$=3.46 Å, $N_2$=3.64 Å, and $CH_4$=3.8 Å) (*2*). This indicates that the GREA models can capture the critical role of diffusion in the gas permeation process, which applies to the two candidates belonging to the polyimide class. However, it is important to note that comparing the $R^2$ value of 10 points in this study (two polymers five tasks) with $R^2$ values obtained in previous works or models, where $R^2$ values of around 0.9 can be obtained on the randomly selected holdout test datasets (usually 20% of the entire dataset) (*28–30*), is not meaningful. The sample sizes differ significantly, making such comparisons invalid.

We find that using unlabeled data is important in improving the accuracy of these extrapolative tasks. As mentioned earlier, the performance of a model herein is characterized by whether the model can correctly predict above or below the upper bounds for five different gas pairs of the two synthesized polymers. If we use GREA trained only using labeled data (GREA-L), with no SGIR algorithm implemented, the accuracy becomes 5/10 (see SI, Fig. S2). While the validation error of the GREA-L is smaller than that of the GREA (see SI, Table S1), its extrapolative predictivity is worse. This may be attributed to the overfitting of GREA-L given the way we trained these models using all the labeled data for training and part of it for validation. The GREA model predictions are also more accurate than the conventional GNNs without environmental augmentation, both trained using labeled and unlabeled data (with SGIR) and only using labeled data (without SGIR, denoted by "-L"), *e.g.*, for GCN, accuracy=5/10; for GIN, accuracy=5/10; for GCN-L, accuracy=5/10; and for GIN-L, accuracy=6/10 (see SI, Figs. S3-S6). This highlights the benefit of promoting the rationale features in the model training, which is helpful especially given the small training data. In the meantime, the GNN-based models are apparently superior to non-GNN-based models regarding the validation MAE (see SI, Table S1), as well as the prediction accuracy of the two tested polymers (RF, accuracy=4/10, GPR, accuracy=5/10, and MLP, accuracy=6/10; see SI, Figs. S7-S9), suggesting the advantage of graph representation learning over traditional molecular fingerprints for polymer representation. Instead of hand-crafted or static fingerprint features, graph representation learning can automatically extract the important features through learning for different downstream tasks. The GREA model is much better than GPR used in previous gas permeability ML studies, but it is noted that the GPR model trained in Ref. (*29*) has some of the labeled data different from the dataset used in the present study, which we do not have access to. Ref. (*46*)'s model is integrated into Polymer Genome (*47*), which is by far the only permeability model that is openly accessible. We used the Polymer Genome platform to predict the gas permeabilities of our experimental polymers, but the predicted selectivity-permeability positions for the different gas pairs are way below the upper bounds (square symbols in Fig. 3b-f), which are also far from the actual experimental results. Similar prediction performance (see SI, Fig. S10) is also observed for the pre-trained model with the best performance on the test data (model: "DNN_BLR_fing"; accessed through: https://github.com/jsunn-y/PolymerGasMembraneML)

reported in Ref. (*30*), suggesting the efficiency and accuracy of our proposed framework dealing with small and imbalanced data.

**Performance Analysis of the Two Polymers**

For P130093, its performance is above the upper bounds of the $O_2/N_2$, $H_2/CH_4$, and $H_2/N_2$ gas pairs, while for P432352, its performance is above the upper bounds of the $H_2/CH_4$ and $H_2/N_2$ gas pairs. Commercial gas separation membranes, *e.g.*, Matrimid®, CA-2.45, and PSF, mostly have high gas permeability and selectivity while maintaining good physical and mechanical properties (*2*). As shown in Table 2 and Fig. 3b-f, our newly identified P130093 outperforms the commercial polymers by a large margin for $O_2/N_2$, $H_2/CH_4$, and $H_2/N_2$ gas pairs, and it also has higher selectivities than commercial ones for $CO_2/CH_4$ and $CO_2/N_2$ separations. For P432352, its performance is much better than commercial ones for $H_2/CH_4$ and $H_2/N_2$ pairs with the rest gas pairs comparable to them. Among these, $O_2/N_2$ separation is well-known to be difficult because of the relatively small kinetic diameter difference between the oxygen molecule (kinetic diameter = 3.46 Å) and nitrogen molecule (3.64 Å). The fact that we found a polyimide polymer guided by the GREA model to have $O_2/N_2$ separation performance above the upper bound itself is of significant practical importance. It is worth noting that there are existing polymers with high permeabilities above the upper bound, but almost all of them have smaller selectivity than the P130093 polyimide (see SI, Fig. S11a). These polymers are mainly based on polymers with high free volume, which are at the laboratory research stage, expensive, and difficult to synthesize. The synthetic accessibility score (SAscore) (*48*) is calculated for each of these polymers and compared with the SAscores of P130093 and P432352 (see SI, Section 3 and Fig. S11b). The SAscore of P130093 is the lowest among the high-performing polymers for $O_2/N_2$ separation, suggesting that the identified P130093 is the easiest to synthesize and thus has greater potential for commercialization. The exceptional $O_2/N_2$ selectivity of P130093 (=17.48) is particularly noteworthy, which is 1.6-6.7× higher than existing gas separation polymers, including PIM-based (exhibits ideal $O_2/N_2$ selectivity in the range of 2.6-6.8), thermally rearranged (exhibits ideal $O_2/N_2$ selectivity in the range of 3.7-8.1), and carbon molecular sieves (CMS)-based (exhibits ideal $O_2/N_2$ selectivity in the range of 2.8-11.1) polymer membranes (*37*). This suggests the power of utilizing ML models to identify those "hidden gems" in the existing polymer database that have simple structures, can be derived from commercially available monomers, and can be processed similarly to commercially available polymer membranes.

**Results Interpretation**

**1. From GREA rationale to force field interpretation**

Between the two identified polymers, P130093 always has higher selectivity while P432352 always has higher permeability regardless of the gas tested (See Fig. 3b-g). We combine the interpretability of the GREA model, semi-empirical analysis, experimental measurements, and MD simulations to comprehensively understand such differences from the molecular level. The rationale separation process in GREA not only helps improve the model performance but provides natural interpretability to the ML prediction, making it no longer an unexplainable black box. The rationale separator predicts the probability of each atom in the polymer graph being classified as the rationales, which are the causal substructures that contribute the most to the downstream property predictions. We emphasize that GREA is self-explainable, which is different from other post-hoc ML interpretation methods based on the rank of feature importance (*e.g.*, SHAP). Oftentimes, the explanation of the feature importance requires predefined hand-crafted structural or chemical information as the input features and some of these predefined features are not directly interpretable with no obvious physical meaning (*30*). However, this is not a problem for the atom-level interpretability from the rationalization of GREA. By visualizing the top 30% rationale subgraphs based on the averaged rationale scores of 10 independent runs, we gain physical insights into the prediction and atom-level molecular features.

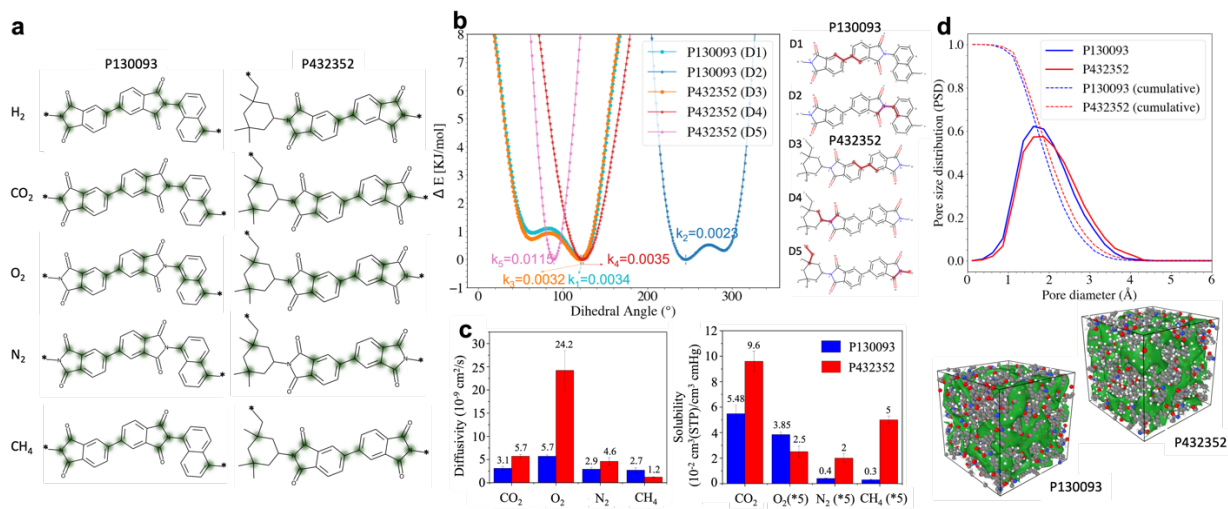

**Figure 4. Result interpretation. (a)** Molecular rationales visualization for P130093 (left column) and P432352 (right column) on five permeability prediction tasks, including $H_2$, $O_2$, $N_2$, $CO_2$, and $CH_4$ (top to bottom). GREA estimates the probability of each node being classified as rationales. The rationale subgraphs are highlighted in green. The darker the color, the higher the rationale score. **(b)** The energy associated with the highlighted dihedral angles of P130093 and P432352. The studied dihedrals are highlighted in red on the right. The force constant k is calculated for each energy well by fitting a quadratic function at the bottom of each well. **(c)** Experimental diffusivity (left) and solubility (right). "(*5)" on the x-axis means the corresponding solubility values are multiplied by five to make the bars look obvious in the graph. $H_2$ is not included in the measurement because it has a very short lag time and cannot be detected by our existing techniques. **(d)** Top figure shows the pore size distribution of the two polymers. The bottom two snapshots show the free-volume elements (colored in green) detected in the MD simulation of the two polymers.

The rationale subgraphs are highlighted in green in the top-30% rationale plots as shown in Fig. 4a. The darker the green color on the atom, the higher the rationale score (also see SI, Fig. S12, the top-50% rationale plot for comparison which includes more atoms). In general, higher attentions are on atoms along the backbone (the shortest path between the two polymerization points, "*"). This makes sense since these atoms and the bonds among the backbone determine its stiffness, which impacts the local segmental motions of polymer chains that further influence the opening and closing of transient free volume elements in the polymer membranes. This phenomenon is captured by GREA. For example, in all cases, the carbon atoms along the backbone which connect two aromatic rings in the dianhydride group are always identified as rationales. These atoms are involved in one of the weakest dihedral angles of the whole structure, twisting the planar structure of the dianhydride group, and weakening chain packing thus yielding a high fractional free volume (FFV) and permeability (see the later semi-empirical analysis for further discussion). Another functional group that is highlighted by GREA is the imide ring group (the ring having a nitrogen atom bonded to two carbonyl groups), which receives higher attention than the aromatic rings in the dianhydride part (the comparison is more obvious in the top-50 % rationale plot; see SI, Fig. S12). This can be attributed to the polar carbonyl groups in the imide ring, which is the key factor in the formation of a charge transfer complex (CTC) between the dianhydride and diamine groups (*49*). The CTC in polyimides is known to increase the interchain attractive forces thus effectively increasing the chain rigidity and enhancing the selectivity. The two polymers have different diamine parts, and the model can also capture the difference. For example, the naphthalene group in P130093 is highlighted by the aromatic carbon atoms along the backbone, which are directly related to the dihedral angles that control the free rotation of naphthalene moiety around the imide linkages (*e.g.*, C-N bonds), whereas the cyclohexane ring structure in P432352 is mostly accentuated by the two carbon atoms with pendant groups (*e.g.*, the methyl group). The nonplanar structure of the cyclohexane ring as well as the methyl groups adds to the steric frustration between polymer chains, resulting in more free-volume elements in the system, therefore, contributing to the high permeability of P432352.

We used a semi-empirical force field to further analyze the strengths of dihedral angles in the backbones identified by the GREA rationale (see Method for semi-empirical analysis details). The calculated force constants are shown in Fig. 4b, which is a measurement of the stiffness of the corresponding dihedral angle. Dihedral angles D1 and D3 (see Fig. 4b), which are the identical structures connecting two aromatic rings in the dianhydride group in both polymers, have almost the same shape of energy well and force constant ($k_1$=0.0034, $k_3$=0.0032). Given that both polymers have the same dianhydride group, different diamine groups of P130093 and P432352 cannot change the stiffness of this common dihedral angle located inside the dianhydride part. As for the difference between P130093 and P432352, the strengths of dihedral angles D2, D4 and D5 are examined. Only one dihedral angle (D2) is calculated for P130093 due to its structural symmetry, whereas two dihedral angles (D4 and D5) are considered for P432352. The energy constants of the two dihedral angles in P432352 ($k_4$=0.0035, $k_5$=0.0115) are larger than the energy constant of the dihedral angle in P130093 ($k_2$=0.0023), indicating the higher stiffness of the backbone in P432352. A more rigid structure often leads to a less efficient chain packing (*7*), offering more free-volume elements for the diffusion of gas molecules. This again interprets why P432352 has higher gas permeabilities. Such backbone stiffness will influence the packing of the polymers and thus the FFV, which is critical to the permeability and selectivity. This semi-empirical force field analysis also in turn proves the effectiveness of the self-explaining ability of GREA.

## 2. Gas transport properties analysis

To further understand the high performance of the two polymers, we measured their glass transition temperature ($T_g$), and calculated their FFVs, diffusivities, and solubilities (see Method section for measurement and calculation details). As shown in Table 3, a high $T_g$ of 300 °C (for P432352) and a non-detectable $T_g$ (for P130093) imply the high rigidity of the polymer backbones, which is beneficial for sieving gases based on their sizes. Moreover, both polymers have moderate to slightly high FFV, which corroborates their high performance. Besides, P432352 has a slightly higher FFV of 16.0% than P130093 (14.5%), which explains why P432352 has higher gas permeabilities. From Fig. 4c, we can see the tested light gases such as $O_2$, $N_2$, and $CH_4$ displayed size-sieving characteristics for both polymers, evidenced by the direct proportion between their diffusivities corresponding to their kinetic diameters. Moreover, both polymers show strong adsorption selectivity for $CO_2$ with high solubilities, which is a typical observation for most polyimides.

To support the experimental results, MD simulations were used to study these two polymers and calculated their FFVs and pore size distributions (PSD) (see Method section for simulation details). The ability of our MD simulations to correctly predict the trend of FFV is confirmed by comparing the results with the experimental values in the literature for different polyimides (*50*) (see SI, Fig. S13 and Table S2). As shown in Table 3, the FFVs from MD simulations show the same trend as the experimentally calculated FFVs, where P432352 possesses a slightly higher fraction of free volumes than P130093, resulting in higher gas permeabilities. The PSD plot in Fig. 4d also supports this finding, where the PSD of P432352 is right-shifted relative to that of P130093, suggesting that P432352 generally has larger pores, which contribute to higher gas permeabilities but lower selectivities than P130093.

Table 3. Experimentally measured and calculated properties (density, FFV and $T_g$) and MD calculated FFV (denoted as FFV-MD) of the two synthesized polymers.

|  | P130093 | P432352 |
|---|---|---|
| Density (g/cm³) | 1.3119 | 1.1652 |
| FFV | 14.5% | 16.0% |
| $T_g$ (°C) | Not detectable | 300 |
| FFV-MD | 15.6% | 17.7% |

**DISCUSSION**

In this section, we further discuss several important observations for the different ML models used. Without using unlabeled data to balance the data distribution, the prediction performance of GREA-L against that of the baselines (GCN-L, GIN-L, RF, GPR, and MLP) is compared in Table S1. Significant improvements in prediction performance over all baselines and prediction tasks are observed from GREA-L. Instead of a hand-crafted or static fingerprint feature, graph representation learning can automatically extract the important features through learning for different downstream tasks. Moreover, the realization of rationale separation and environment replacement augmentation is the key to the success of GREA-L. It allows the model to focus only on the rationale part of the input polymer graph and be robust to the noise to maximize the performance of the downstream task.

However, to prevent the models from overfitting on the validation set since it is also part of the training data, unlabeled data is introduced. We examine the performance of the GNN-based models using both the labeled and unlabeled data (GREA, GCN, and GIN). Compared with GNN-based models trained only on the labeled data (GREA-L, GCN-L, and GIN-L), leveraging the unlabeled data during training iterations can undermine the model performance on the validation set, *i.e.*, all the GNN-based models see higher average MAE after adding the unlabeled data (see Table 1 and Table S1). This can be attributed to the mitigation of overfitting on the validation set and the gain of model generalizability at the expense of validation accuracy. In addition, GREA is still the best among the models leveraging the unlabeled data, except for one case of $CO_2$ permeability prediction, again suggesting the improved performance from graph rationalization and environment replacement augmentation.

It is worth noticing that two model performance metrics were used to evaluate and compare different models: the MAE on the validation set (above-the-bound training data) and the binary classification accuracy on the two test polymers. The validation set is used to select the best model, i.e., the GREA model, since the test accuracy is unknown before synthesizing and experimentally measuring the polymer properties. The test accuracy serves as a posterior verification of our method. The variations in validation and test performance between the models, such as MLP outperforming GCN-L in the test data (see SI, Fig. S4 and S9) but underperforming in validation data (see SI, Table S1), can be potentially attributed to the problem of overfitting. GCN-L could potentially be more prone to overfitting than MLP since it has a larger number of parameters (refer to the Methods section for detailed model structures).

The focus of this study is to build up efficient and effective ML models to identify high-performance polymer membranes in the existing PoLyInfo database that are suitable for industrial scaling up. While there are state-of-the-art polymers that have significantly outperformed the 2008 upper-bound for gas separation, we did not use the most recent upper-bound lines in our study. This is because the majority of the polymers that define the latest upper-bound lines are highly complicated polymers with ladder or semi-ladder backbone structures, that involve highly complex synthesis procedures (*2*, *9*, *10*, *51*). These complex polymers are not well represented in the PoLyInfo database, which is intended to serve as a more general application database for polymers. Instead, we aim to identify "hidden gems" in the PoLyInfo database that have accessible structures, preferably can be derived from commercially available monomers, and can be processed similarly to commercially available polymer membranes. Our potential next step is to use ML techniques to design more complicated polymer structures, such as ladder polymers, for improved gas separation performance. Nevertheless, our discovered polymers, although much simpler than these more complicated polymers, showed $O_2/N_2$ selectivity that is 1.6-6.7× higher than existing gas separation polymers, including ladder and thermally rearranged polymers (*37*).

**CONCLUSION**

In conclusion, we have demonstrated the ability of graph augmented ML model, GREA, and graph imbalanced ML framework, SGIR, to guide the design of polymers with performances simultaneously exceeding the theoretical upper bounds in separating multiple gas pairs. The two predicted promising candidates that are experimentally synthesized and tested are able to validate the ML model prediction. The exceptional gas separation performance of the two polymers, especially the ultrahigh $O_2/N_2$ selectivity of P130093, highlights the potential of utilizing ML models to accurately identify simple-structured polymers from commercially available monomers in the existing polymer database. The rationales in the GREA model can shed light on the molecular-level origin of the high performance of these polymers. This ML model interpretation is further supported by the experimental characterization and calculation of the FFV and $T_g$, the semi-empirical forcefield analyses of key dihedral angles identified by GREA, and the MD calculation of FFV and PSD. Our study presents a useful ML technique to tackle small and imbalanced data problems, which may be generalized to other materials development tasks, and the developed polymers may be used for high-performance membranes to separate several industrially important gas pairs.

**METHODS**

**Polymer Representations**

One of the key challenges in polymer informatics (PI) is to choose appropriate numerical descriptors for polymer structures that can serve as feature vectors for downstream ML applications. Among the descriptors commonly used in PI are molecular fingerprints and molecular graphs. Regarding molecular fingerprints, the two types of most widely used methods are structural keys and hashed fingerprints. Structural keys encode the structural information of a molecule into a binary bit vector, and each bit position corresponds to whether the molecule possesses a predefined structural feature. A representative structural key is the MACCS key, containing 166 substructures or fragments (*52*). In contrast to structural keys, hashed fingerprints do not require a library of predefined building fragments and are created by enumerating all the possible fragments up to a given size and converting the fragments into numerical values based on a hash function. Extended-connectivity fingerprints (ECFPs) (*53*) is the most popular hashed fingerprinting method considering the circular atom neighborhood. Besides molecular fingerprints, a more natural representation of organic molecules like polymers is molecular graphs, in which atoms are represented by nodes and bonds by edges. Molecular graphs are usually further described by three matrices: the node feature matrix, the edge feature matrix, and the adjacency matrix. Using graphs to represent molecules is advantageous since it can be directly fed into GNNs for representation learning with rich structural information, feasible for both molecular property prediction and inverse design. The molecular graph property prediction task generally consists of two phases: a graph encoder to learn a fixed-length molecular representation from the graph and a prediction decoder to map the learned representation to the target property. The GNN-based representation learning (encoder) consists of three general operations: message passing, node update, and readout. The message passing module updates the node representations iteratively with aggregated information from the neighborhood. After getting the final node representation, the graph-level representation is computed through a readout function. Prevalent GNN models used in molecular graph representation learning are GCN, GIN, etc.

In this work, we use both molecular fingerprints and polymer graph representation from GNNs to represent polymers. The learned latent spaces of five different tasks ($H_2$, $CO_2$, $O_2$, $N_2$, and $CH_4$) for the training and screening datasets in this work are visualized using t-distributed stochastic neighbor embedding (t-SNE) (*54*) in a two-dimensional map in SI, Fig. S14. The PoLyInfo screening dataset generally spans the feature space of the training data of the five tasks. Details of molecular fingerprint and graph representation of this work are as follows:

*1. Molecular fingerprint:* Both the MACCS keys and ECFPs with a radius equals to 2 (namely, ECFP4) are generated through RDKit (*55*). The molecular fingerprints play two roles: (1) serving as the polymer representation input for non-GNN-based ML models; (2) being concatenated with graph embedding in the latent space for GNN-based ML models to enrich chemical knowledge captured by the latent space and regularize the representation learned by the graph encoder.

*2. Graph representation:* SMILES files of polymers are converted into polymer graphs for further GNN-based computation. RDKit (*55*) is utilized to extract node and edge features. Nine categories for node features are defined, including (1) atomic number; (2) chirality; (3) degree of the atom; (4) formal charge; (5) total number of Hs (explicit and implicit) on the atom; (6) number of radical electrons; (7) hybridization; (8) aromaticity; (9) is in a ring or not. Three categories for edge features are defined, including (1) bond type; (2) stereo configuration; (3) is conjugated or not. Then each polymer is treated as an undirected graph $G = (X, E, A)$, where $X$ is the node feature matrix, $E$ is the edge feature matrix, and $A$ is the adjacency matrix. Assuming there are $n$ nodes and $e$ edges in the graph, $d$ and $b$ are the number of features for node and edge respectively, then $X \in \{0,1\}^{n \times d}$, $E \in \{0,1\}^{e \times b}$, and $A \in \{0,1\}^{n \times n}$. All these features are then mapped into a continuous latent space using trainable parameters, and the representation is optimized during the model training process.

**Training of ML Models**

All labeled data are used for training for any of the specific gas permeability prediction tasks. The validation set is deliberately selected to be the polymers that are above any of the five upper bounds, which helps improve the model prediction accuracy in the sparsely populated target regime. The GREA model has a two-layer GIN as the rationale separator, a five-layer GIN as the graph encoder, and a three-layer MLP as the property predictor. It is trained with a batch size of 128, the Adam optimizer regularized with a 0.01 weight decay, the initial learning rate set as 0.001 with a cosine scheduler, the dropout rate of 0.5, and maximal training epochs of 600 with early stopping. The dimension of the latent space is set as 600. The baseline GCN and GIN models have the same architecture as the GREA model's graph encoder and prediction decoder. As for non-GNN baselines, the random forest (RF) regressor is implemented using 100 estimators, with mean squared error as the split quality measurement and the minimum number of samples required to split a node as 30. The GPR is implemented with a radial basis function kernel of a length scale set as 1.0 and lower and upper bound set as 0.01 and 100, respectively. Alpha is set as 5, which is the value added to the diagonal of the kernel matrix during fitting to ensure a positive definite matrix. The MLP regressor is implemented with 1 hidden layer of 600 neurons, a batch size of 128, the ReLU activation function, the stochastic gradient descent optimizer, and maximal training epochs of 200 with early stopping. All methods run 10 times independently.

As mentioned in the Results section, the rationale separator divides a polymer graph into two subgraphs, a rationale subgraph and an environment (noise) subgraph. It is important to clarify that the term "noise" used here does not refer to "noisy data", which are data points that are irrelevant or unrepresentative of the underlying patterns in the data and should identified, removed or minimized (*56*). Rather, "noise" is used here to describe small perturbations in the polymer structures (other examples include Gaussian noise), which is utilized by many data augmentation methods in ML to create new data points. The environment subgraph in GREA serves as a type of noise learned from the data. It fits the goal of data augmentation and can be used to create numerous and diverse labeled training data to enrich the dataset and overcome the problem of limited supervision.

A semi-supervised framework for graph imbalanced regression (SGIR) (*36*) is embedded with GREA, GIN, and GCN to address the data insufficiency and label imbalance problems. It is a self-training algorithm that is capable of gradually reducing the model bias resulting from data imbalance by iteratively producing high-quality training data in the rare label value region. Specifically, SGIR first defines a novel confidence score using environmental subgraphs, selects highly confident polymers from the predicted labels of the large unlabeled polymer dataset PoLyInfo, and removes noisy unlabeled polymers. Then in the latent representation space, it further augments polymers in label areas that seriously lack data by a label-anchored mixup algorithm. Details of the SGIR algorithm can be found in Ref. (*36*). After training the model for 60 epochs, we start to implement the SGIR algorithm every 30 epochs. Since SGIR is not applicable to non-GNN-based models, the RF, GPR, and MLP models are trained in a supervised manner.

All methods are implemented using Pytorch, Scikit-learn, and Pytorch Geometric on Linux with an Intel Xeon Gold 6130 Processor (16 Cores @2.1Ghz), 96 GB of RAM, and an RTX 2080Ti card (11 GB RAM). The code for our ML implementation can be found at https://github.com/Jiaxin-Xu/PolyGasPerm-GREA.

**Experimental Synthesis and Permeation Test**

*1. Synthesis and film casting of polymer P130093:* Polymer P130093 (*57*) was synthesized by dissolving 0.8030 g (5.076 mmol) of 1,5-diaminonaphthalene in 4 ml anhydrous NMP (N-Methyl-2-pyrrolidone) at 80 °C. This was followed by the addition of an equimolar amount of 3,3',4,4'-biphenyltetracarboxylic dianhydride and 11 ml anhydrous NMP to maintain 15 wt.% solid content while the temperature was maintained until complete dissolution of both monomers. Then the reaction continued at room temperature for another 4 hours to obtain a viscous polyamic acid (PAA) solution. To complete the imidization and obtain thin films of the final polyimide, the PAA solution was diluted to 7.5 % concentration with anhydrous NMP, filtered with 0.45 μm Teflon filters, and cast on glass plates under an infrared lamp at ~ 60 °C for 24 h. It was then dried at 150 °C under vacuum for 12 h, soaked in methanol for 3 h and dried again at 150 °C under vacuum for 12 h. The solvent-free PAA film was sandwiched between two porous ceramic plates and thermally imidized in a muffle furnace under nitrogen flow, where the temperature was ramped at 10 °C per minute to 180, 210, 250, 350, and 400 °C, maintaining 15 minutes at each temperature before finally cooling to room temperature at no greater than 10 °C per minute. The fully imidized structure of the solvent-free films was confirmed by proton nuclear magnetic resonance spectroscopy ($^1$H NMR) and Fourier transform infrared spectroscopy in attenuated total reflectance mode (ATR-FTIR).

*2. Synthesis and film casting of polymer P432352:* Polymer P432352 (*58*) was synthesized by reacting 1.7 g (9.98 mmol) of 5-amino-1,3,3-trimethyl cyclohexane methylamine with a stoichiometric amount of 3,3',4,4'-biphenyltetracarboxylic dianhydride in 42 ml *m*-cresol at 11.2 wt.% concentration in a flame-dried 3-neck flask fitted with a mechanical stirrer. The monomers were fully dissolved within 1 h at 70 °C and then maintained at 90 °C for 3 h. Then the temperature was gradually raised to 200 °C within 2 h before 10 ml ortho-dichlorobenzene was added, and a Dean-Stark trap was attached for azeotropic reflux to complete the imidization over another 4 h at 200 °C. Fiber chunks of the product were obtained by precipitating the viscous solution into 600 ml stirring methanol, which was collected via filtration and dried at 100 °C for 12 h in vacuo. The thin film of the polymer was obtained by solution casting on a circular glass plate using chloroform as the casting solvent. 1.6 % w/v of the polymer solution was cast at room temperature under nitrogen flow for 48-72 h enabling slow evaporation of the solvent.

*3. Gas permeation tests:* The pure permeabilities of the polymers for $H_2$, $CH_4$, $N_2$, $O_2$, and $CO_2$ were measured at 35 °C using a constant-volume variable-pressure method (*45*). Thin films (44-60 μm in thickness) of the polymers were mounted on aluminum duct tape with the aid of epoxy glue and protected on the backside with filter paper. The exposed film region was scanned with ImageJ to measure the available area for gas permeation and then loaded into the gas cell with the sampler holder immersed in the deionized water bath for temperature control. The entire system (upstream and downstream sides) was degassed in vacuo for at least 12 h before introducing ultra-high purity grade gases that were maintained at 30, 50, and 80 psig until a steady state increase in pressure vs time in the downstream was achieved. The permeability for each gas was calculated using the Eq. (1):

$$P = 10^{10} \frac{V_d l}{P_{up} TRA} \left[ \left(\frac{dp}{dt}\right)_{ss} - \left(\frac{dp}{dt}\right)_{leak} \right],$$

*Eq. (1)*

where $P$ (Barrer, 1 Barrer = $1 \times 10^{-10} cm^3$ [STP] $cm^2/(cm^3$ s cmHg)) is the gas permeability, $l$ is the film thickness (cm), $V_d$ is the calibrated downstream volume ($cm^3$), $P_{up}$ is the upstream pressure (cmHg), $A$ is the effective film area ($cm^2$), $\left(\frac{dp}{dt}\right)_{ss}$ and $\left(\frac{dp}{dt}\right)_{leak}$ are the steady-state pressure increment in downstream, and the leak rate of the system (cmHg/s), respectively; $T$ is the test temperature (K), and $R$ is the gas constant (0.278 $cm^3$ cmHg/($cm^3$ (STP) K)). Details of the gas permeation test are summarized in SI Table S3 and S4. The ideal selectivity ($\alpha_{A/B}$) for two different gases A (more permeable) and B is defined as the ratio of pure gas permeability of the two gases and is calculated as Eq. (2):

$$\alpha_{A/B} = \frac{P_A}{P_B}.$$

*Eq. ( 2 )*

**Dihedral Angle Analysis**

Gaussian 16 is used to study the conformational flexibility of P130093 and P432352 by calculating the energy change associated with the change in the dihedral angles of interest. The initial polymer structures consisting of two repeated units (each the polymerization point at both ends is replaced by a hydrogen atom) are first built in Gaussview with energy minimization. Then, a relaxed potential energy scan is conducted where the specific dihedral angle of interest is fixed in each scan, and other parts of the molecule are relaxed to calculate the total energy. The semi-empirical method (pm6) is adopted throughout the simulations.

**Experimental Measurement and Calculation of Diffusivity, Solubility, Density, FFV, and $T_g$**

The averaged-diffusion coefficient, D ($cm^2s^{-1}$), was calculated using the lag-time method. Details of calculation are summarized in SI Table S3 and S4. The solubility coefficient, S ($cm^3$ (STP)/$cm^3$ atm), was obtained using the relationship, S = P/D. The $T_g$ of the polymers is determined using the Differential Scanning Calorimetry Q2000 by TA Instruments with 50 mL/min nitrogen purge at 10 °C/min heating rate during the second heating cycle (100-400 °C). The polymer densities are obtained by the buoyancy technique, which relies on Archimedes' principle. Thin films of the polymers are weighed in dry and wet forms using an analytical balance ML 204 by Mettler Toledo fitted with density measurement kit deionized water at room temperature. The FFVs are computed using Bondi's group contribution method (*26, 59*).

**MD Simulation Analysis of Membrane Pores**

The procedure of calculating FFV and PSD for the two polymers using MD simulations is composed of two steps: amorphous polymer structure generation and pore structure characterization. Details are shown below.

*Step 1: Amorphous polymer structure generation:* Taking the SMILES of the polymer as an input, the initial amorphous polymer structure is generated by a Python pipeline based on PYSIMM (*60*). It generates a polymer chain through polymerization, with the number of atoms per chain fixed to around 600. Then the chain is replicated, and a system of six chains in total is generated and enclosed in a simulation box. Meanwhile, the GAFF2 (General AMBER Force Field 2) (*61*) forcefield parameters are assigned to the polymer system and an input script for MD simulation using the large-scale atomic-molecular massively parallel simulator (LAMMPS) (*62*) is generated. Periodic boundary conditions in all spatial directions are applied. The system is then optimized gradually. First, the system is simulated with electrostatic interactions turned off and Lennard-Jones interactions with a cutoff of 0.300 nm, aiming to eliminate close contact between atoms. An NVT ensemble is applied at 100 K for 2 ps, with a timestep of 0.1 fs, followed by the system heating up from 100 K to 2000 K in 2 ns. Then an NPT ensemble is employed at 2000K and 0.1 atm for 50 ps, after which the pressure is increased from 0.1 to 500 atm in 2 ns with temperature fixed at 2000K. Then the obtained polymer system is directly compressed in all spatial directions so as to match the density measurement from our experiment. After the initialization, the electrostatic interactions are turned on and the PPPM (Particle- Particle- Particle-Mesh)-based Ewald sum method is used. The Lennard-Jones interactions cutoff is set as 1.200 nm. To achieve a reliable amorphous polymer structure, an NVT ensemble is further applied at 2000 K for 0.2 ns, with a timestep of 0.1 fs, before the system is quenched. The snapshots at 0.12, 0.14, 0.16, 0.18 and 0.2 ns of the last step are saved for later pore structure characterization.

*Step 2: Pore structure characterization:* We use PoreBlazer (*63*) to characterize the pore size and distribution, which is calculated based on the Hoshen-Kopelman cluster labeling algorithm. The diameter of the probe is set to 1.25 Å, which is tuned so that it gives a similar magnitude of FFV as the experimentally calculated FFV. The MD-calculated FFV and PSD are averaged across the five different snapshots as described in Step 1.

**SUPPLEMENTAL INFORMATION**

Supplemental Information includes 14 figures and 4 tables.

**Supporting Information for**

Superior Polymeric Gas Separation Membrane Designed by Explainable Graph Machine Learning

Jiaxin Xu[1,+], Agboola Suleiman[2,+], Gang Liu[3,+], Michael Perez[1], Renzheng Zhang[1], Meng Jiang[3,*], Ruilan Guo[2,*], Tengfei Luo[1,2,*]

Affiliations:

[1] Department of Aerospace and Mechanical Engineering, University of Notre Dame, Notre Dame, IN 46556, United States of America

[2] Department of Chemical and Biomolecular Engineering, University of Notre Dame, Notre Dame, IN 46556, United States of America

[3] Department of Computer Science and Engineering, University of Notre Dame, IN 46556, United States of America

[+] Equal contribution

* Corresponding authors: mjiang2@nd.edu; rguo@nd.edu; tluo@nd.edu

**This PDF file includes**:

    Supporting text
    Figures S1 to S14
    Tables S1 to S4
    SI References

**Supporting Information Text**

**Section 1. Machine learning models and performance evaluation**

Several popular baseline models in molecular property prediction, both GNN-based and traditional non-GNN-based, are implemented in comparison with GREA. The graph convolutional network (GCN) is a variant of convolutional neural networks which can operate directly on graphs and encode node features efficiently. The graph isomorphism network (GIN) generalizes the Weisfeiler-Lehman isomorphism test and has high discriminative and representational power among GNNs for graph-level tasks. Besides GNN methods, several traditional ML algorithms are also implemented. Random forest (RF) is an ensemble of decision trees usually trained with the "bagging" method and is one of the most used ML algorithms because of its robustness and simplicity (*1*). Gaussian process regression (GPR) is a probabilistic supervised learning model that can incorporate prior knowledge into its kernel functions and provide predictions along with uncertainty estimations (*2*). Multilayer perceptron (MLP) is a feed-forward neural network model composed of multiple layers of interconnected neurons.

**Section 2. Dataset and data cleaning**

The labeled dataset in this paper is built based on the public Polymer Gas Separation Membrane Database downloaded from the Membrane Society of Australasia (MSA) portal (*3*), which hosts a variety of experimental gas permeability data for around 1500 polymers published from 1950 to 2018. Five major gases in the gas separation industry – nitrogen ($N_2$), oxygen ($O_2$), hydrogen ($H_2$), methane ($CH_4$), and carbon dioxide ($CO_2$) - are mainly included in the measurements. A wide range of membrane materials is covered including rubber and glassy polymers, carbon molecular sieves, and zeolites. To include more up-to-date high-performance gas separation polymers, we manually collected around 90 experimental gas permeability data from 19 studies in the literature published between 2018 and 2022, with a special focus on polyimides. SMILES (Simplified Molecular Input Line Entry System) is a widely used string line notation in the chemistry community for uniquely describing the fundamental chemical components and structures of different molecules (*4*). Polymer-SMILES (p-SMILES) is a specific class of SMILES notation for polymers with "*" denoting the polymerization points of monomers (*5*). For polymer structure representation, we manually added the corresponding p-SMILES strings to every entry of the dataset and removed those entries that (a) cannot be represented as a plain p-SMILES form (e.g., composite materials, doped and modified materials, or materials with no specific repeating units and structures stated in the literature) and (b) are copolymers (herein only homopolymers are considered for the simplicity of polymer representation). Regarding a single polymer structure with multiple experimental values recorded, produced by different synthesizing and testing procedures (e.g., different thermal treatment temperatures, different aging times, etc.), only the median is preserved as a more reliable representation of the polymer gas permeability. Eventually, a total number of 836 homopolymer structures are left in our training dataset, each with at least one permeability record for one of the five major industrial gases. For each gas, the total number of labeled data after cleaning is ($H_2$, 509), ($O_2$, 805), ($N_2$, 794), ($CO_2$, 754), and ($CH_4$, 681). Label distributions for the five gases of interest are shown in Fig. S1. As we can see, the five datasets cover an ideally wide range of gas permeability values, however, with highly imbalanced distributions. This will be detrimental to training a model that can accurately identify high-performance polymer membranes because the target polymers always fall in regions rarely represented by the training data (e.g., two tail regions in the distributions corresponding to either high or low permeabilities).

The large unlabeled dataset is from PoLyInfo (*6*), which is by far the largest experimental polymer database on polymeric materials design. It collects detailed information from literature on chemical structures, synthesizing and processing methods of samples, etc., covering a wide range of polymer subtypes, from homopolymers, copolymers, and polymer blends, to polymeric composites and compounds. In this work, only homopolymer is considered for the present study, which gives us a chemical space of 12,769 candidates for screening with guaranteed synthesizability by learning from the synthesizing procedures reported in the recorded literature.

**Section 3. SAscores calculation**

The synthetic accessibility score (SAscore) (*7*) is calculated for the polymers in the training dataset with above-the-bound $O_2/N_2$ separation performance and compared with the SAscores of P130093 and P432352 (see Fig. S11b). The source code for the calculation of SAscore as described in Ref. (*7*) with several small modifications was adopted and can be found in Ref. (*8*).

**Table S1. Performance of models using only the labeled dataset for the prediction of five gas permeability tasks.** Average MAE values with standard deviations on the validation dataset are shown. The best performance across all models for each task is bolded. All MAE values are in units of $\log_{10}$Barrer, where 1 Barrer = $10^{-10}$cm$^3$(STP)cm/(cm$^2$ s cmHg).

| GNN-based model | Gas | $N_2$ | $O_2$ | $H_2$ | $CH_4$ | $CO_2$ |
|---|---|---|---|---|---|---|
| Yes | GREA-L | **0.448±0.108** | **0.382±0.030** | **0.271±0.033** | **0.402±0.060** | **0.348±0.112** |
| | GCN-L | 0.604±0.255 | 0.434±0.068 | 0.300±0.092 | 0.475±0.071 | 0.374±0.066 |
| | GIN-L | 0.608±0.078 | 0.488±0.088 | 0.359±0.080 | 0.567±0.179 | 0.368±0.066 |
| No | RF | 0.556±0.011 | 0.528±0.010 | 0.493±0.010 | 0.646±0.011 | 0.749±0.013 |
| | GPR | 1.059±0.000 | 1.036±0.000 | 1.084±0.000 | 0.923±0.000 | 1.167±0.000 |
| | MLP | 0.529±0.020 | 0.473±0.022 | 0.558±0.020 | 0.515±0.025 | 0.861±0.027 |

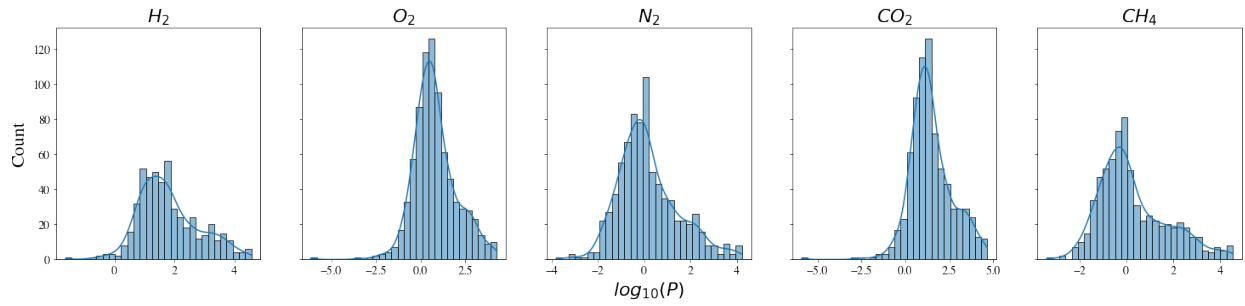

**Figure S1. Label distribution of available permeability data for five major gases: $H_2$, $O_2$, $N_2$, $CO_2$, and $CH_4$.** Permeability (P) is in units of Barrer (1 Barrer = $1\times10^{-10}$ cm$^3$ [STP] cm$^2$/cm$^3$ s cmHg, where STP is the Standard Temperature Pressure, 1 bar and 273 K). The x-axis is in the unit of $\log_{10}$Barrer.

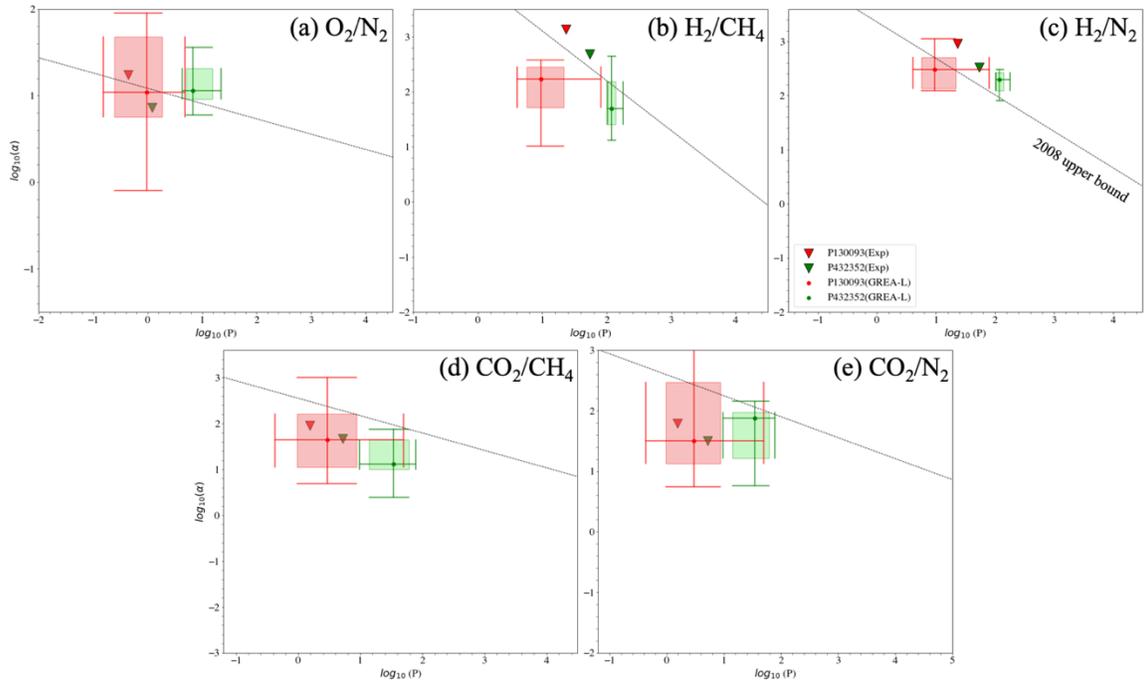

**Figure S2. Robeson plot visualization for five different separation tasks using experiment values and GREA-L predictions.** Accuracy=5/10.

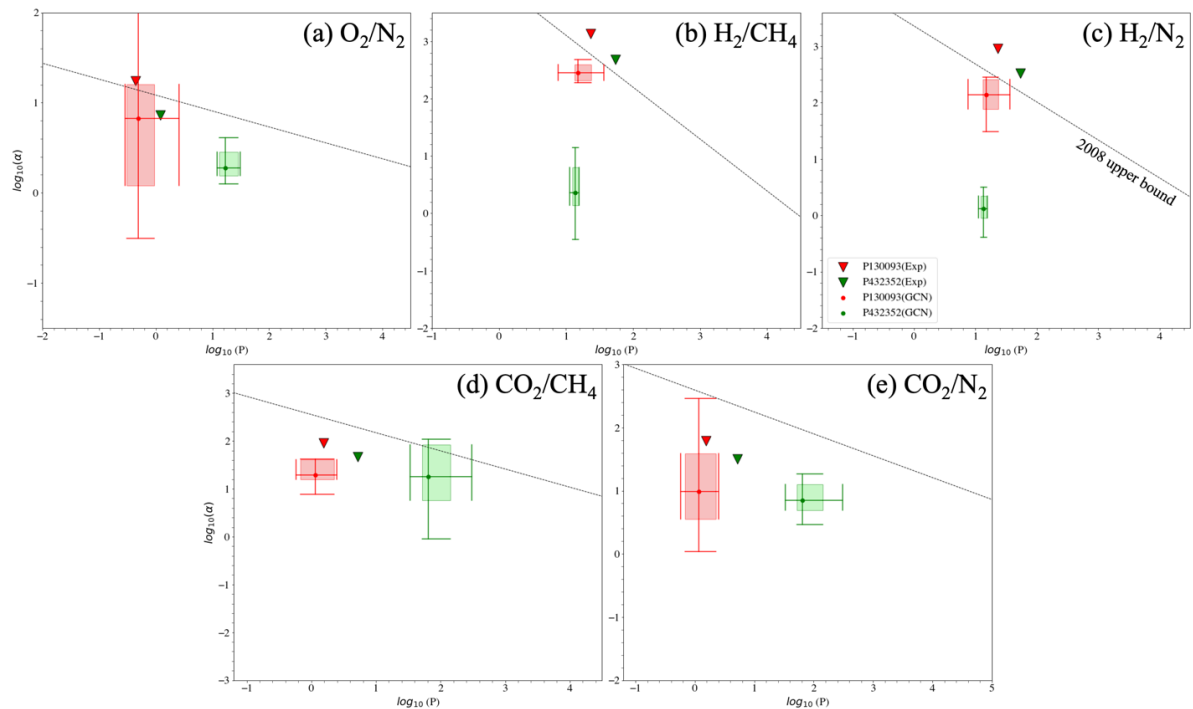

**Figure S3.** Robeson plot visualization for five different separation tasks using experiment values and GCN predictions. Accuracy=5/10.

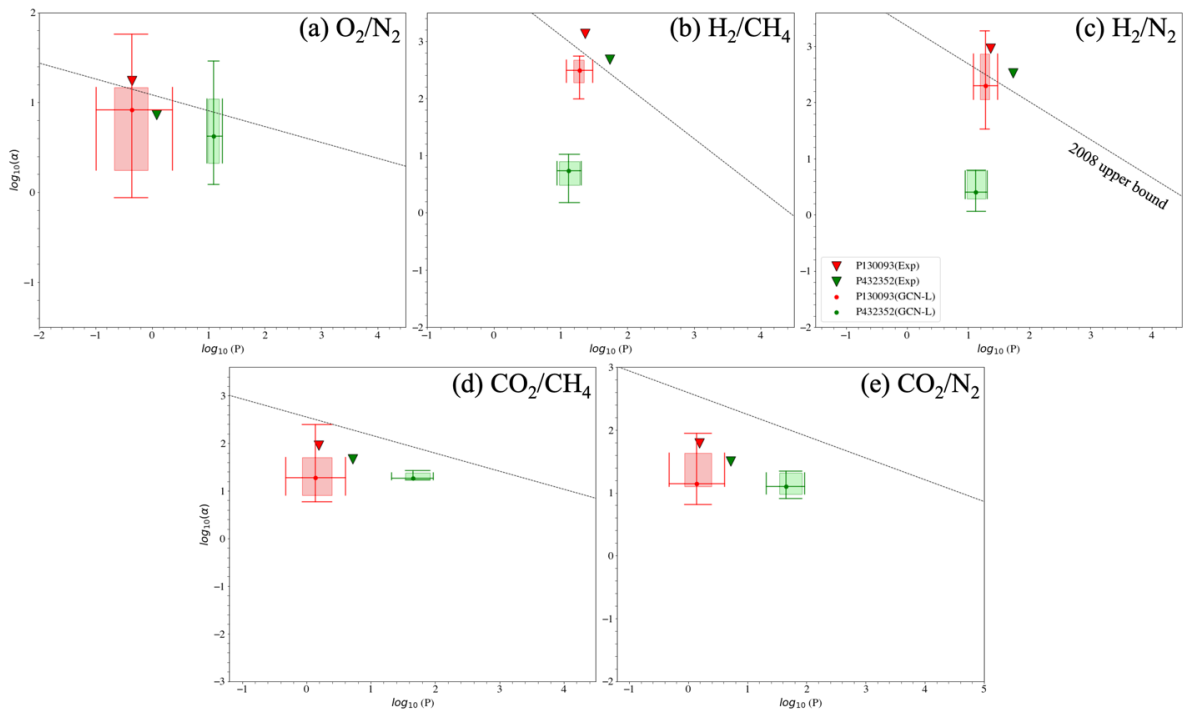

**Figure S4.** Robeson plot visualization for five different separation tasks using experiment values and GCN-L predictions. Accuracy=5/10.

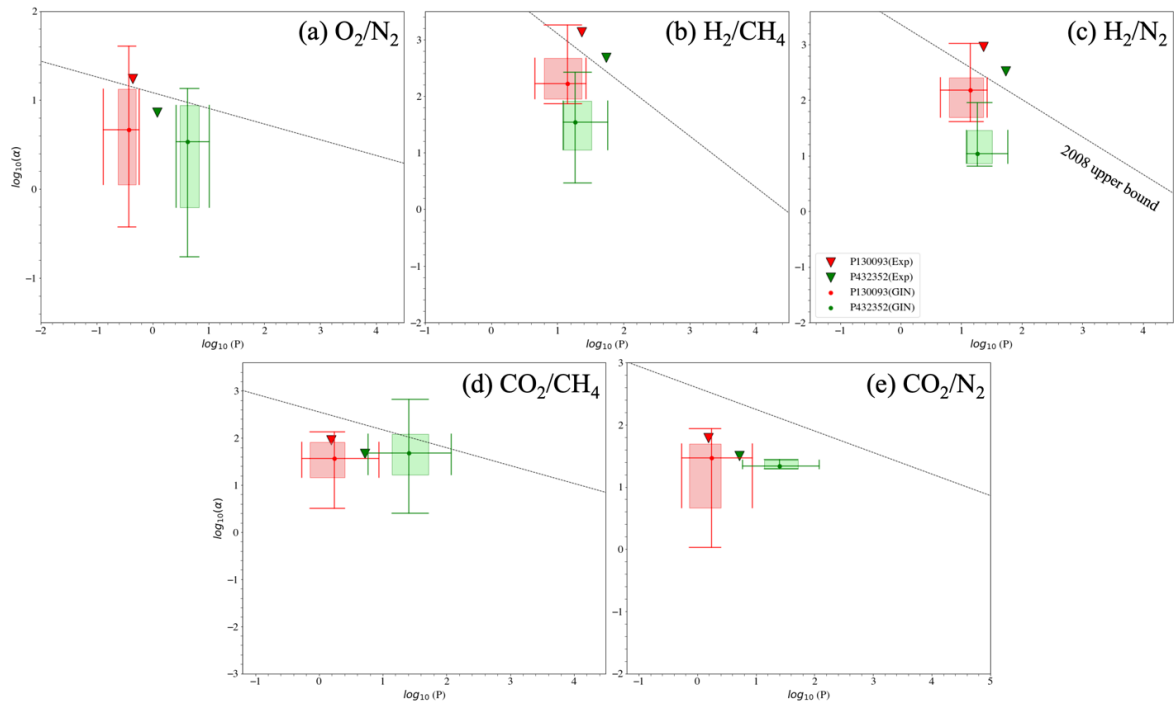

**Figure S5.** Robeson plot visualization for five different separation tasks using experiment values and GIN predictions. Accuracy=5/10.

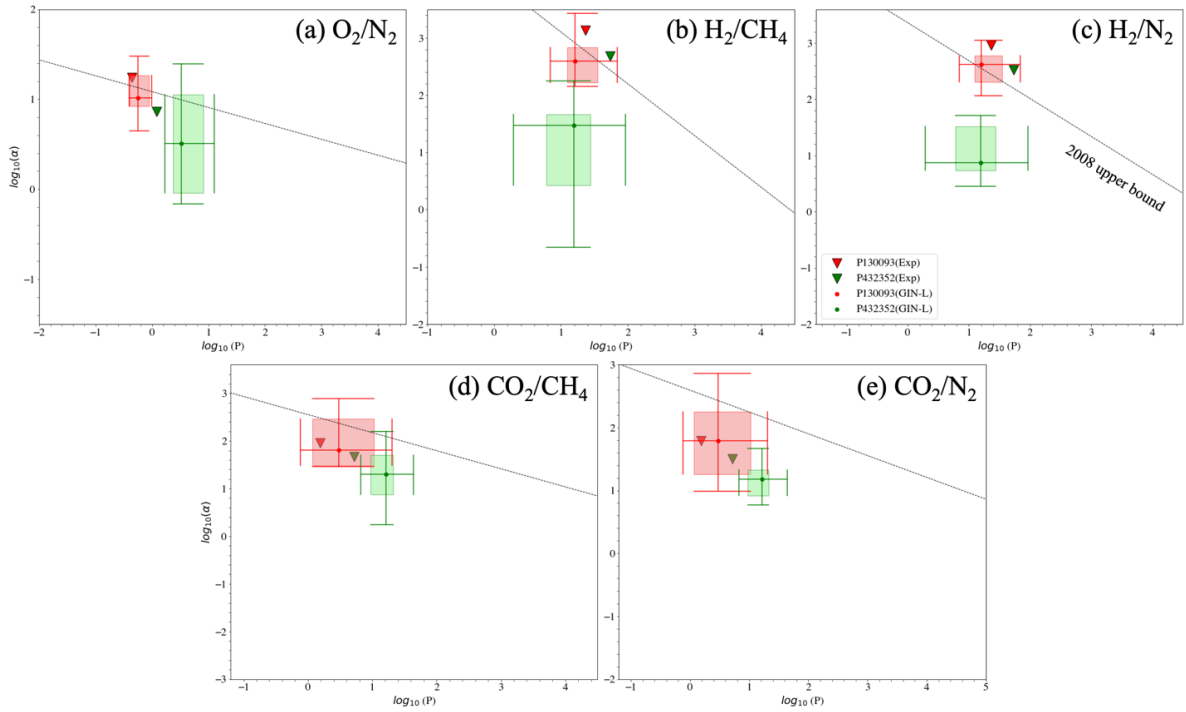

**Figure S6.** Robeson plot visualization for five different separation tasks using experiment values and GIN-L predictions. Accuracy=6/10.

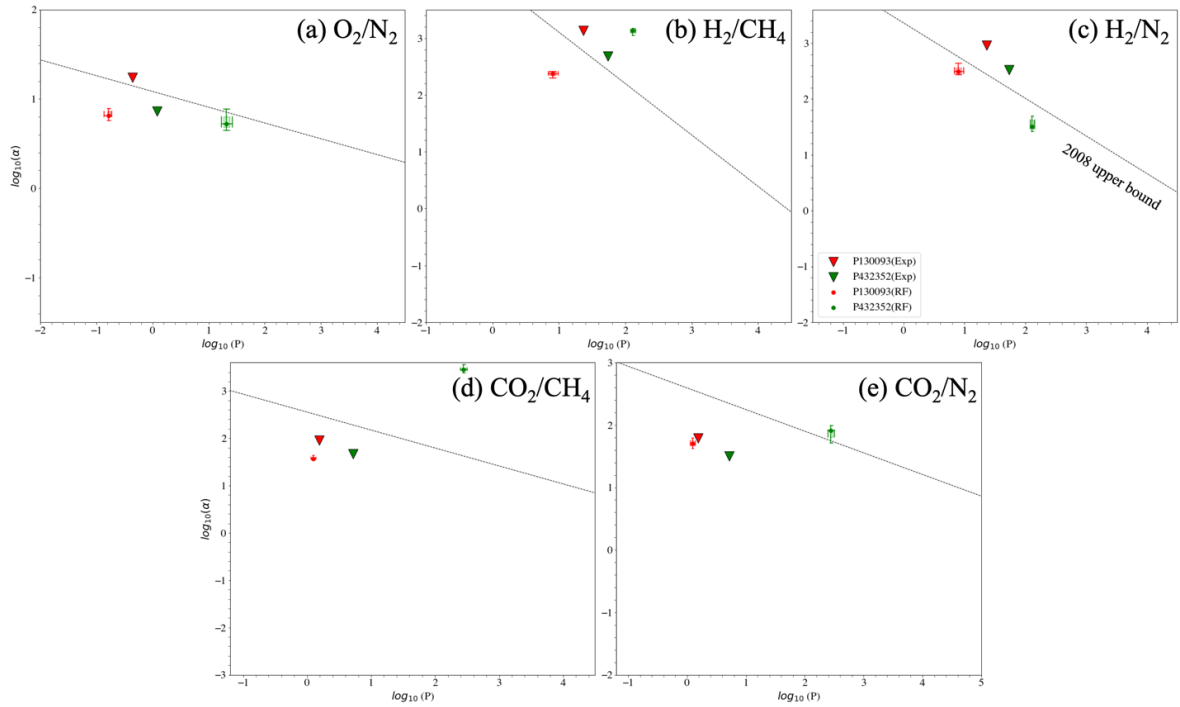

**Figure S7.** Robeson plot visualization for five different separation tasks using experiment values and RF predictions. Accuracy=4/10.

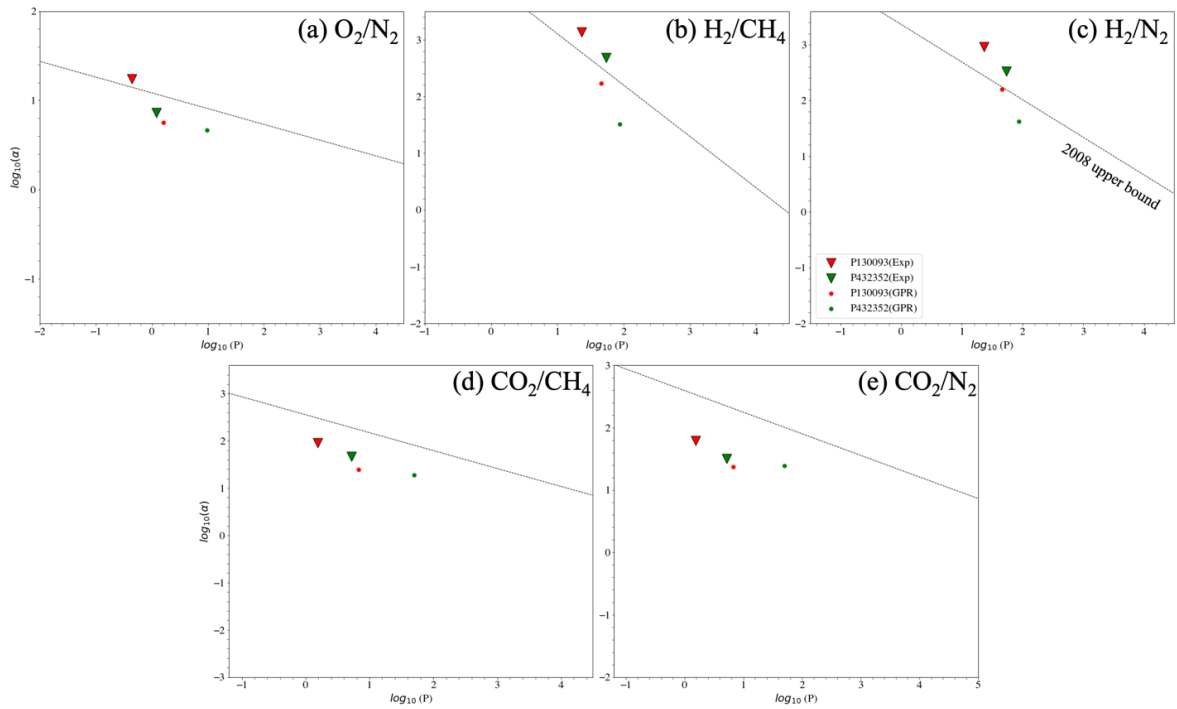

Figure S8. Robeson plot visualization for five different separation tasks using experiment values and GPR predictions. Accuracy=5/10.

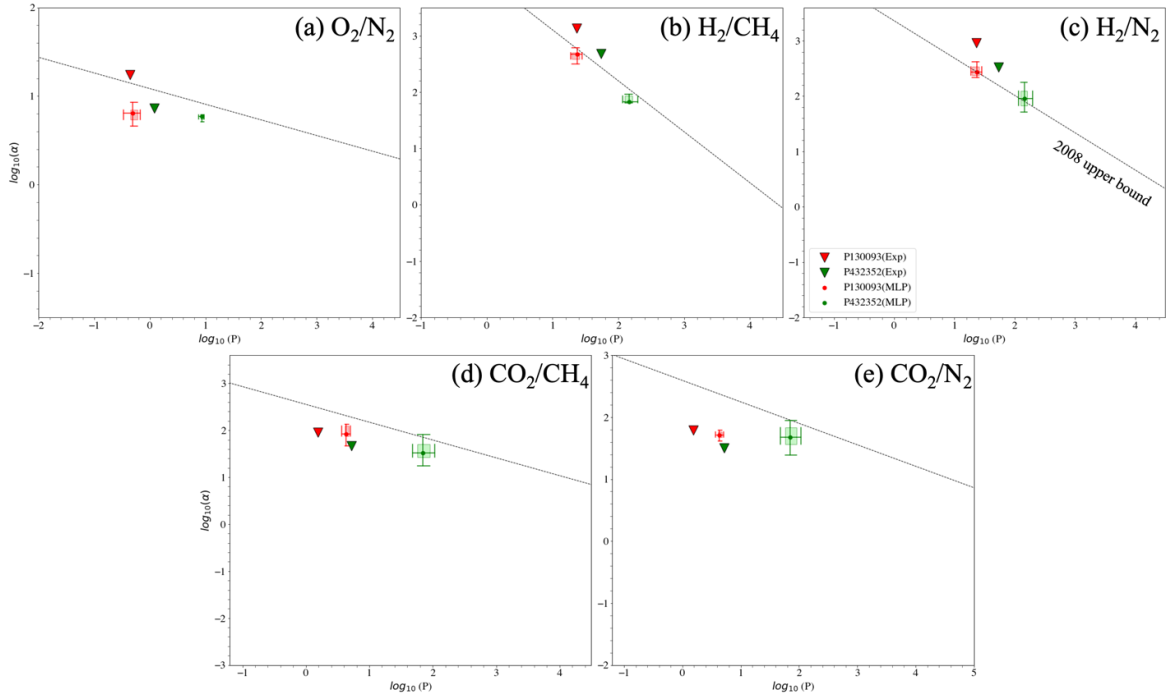

**Figure S9. Robeson plot visualization for five different separation tasks using experiment values and MLP predictions.** Accuracy=6/10.

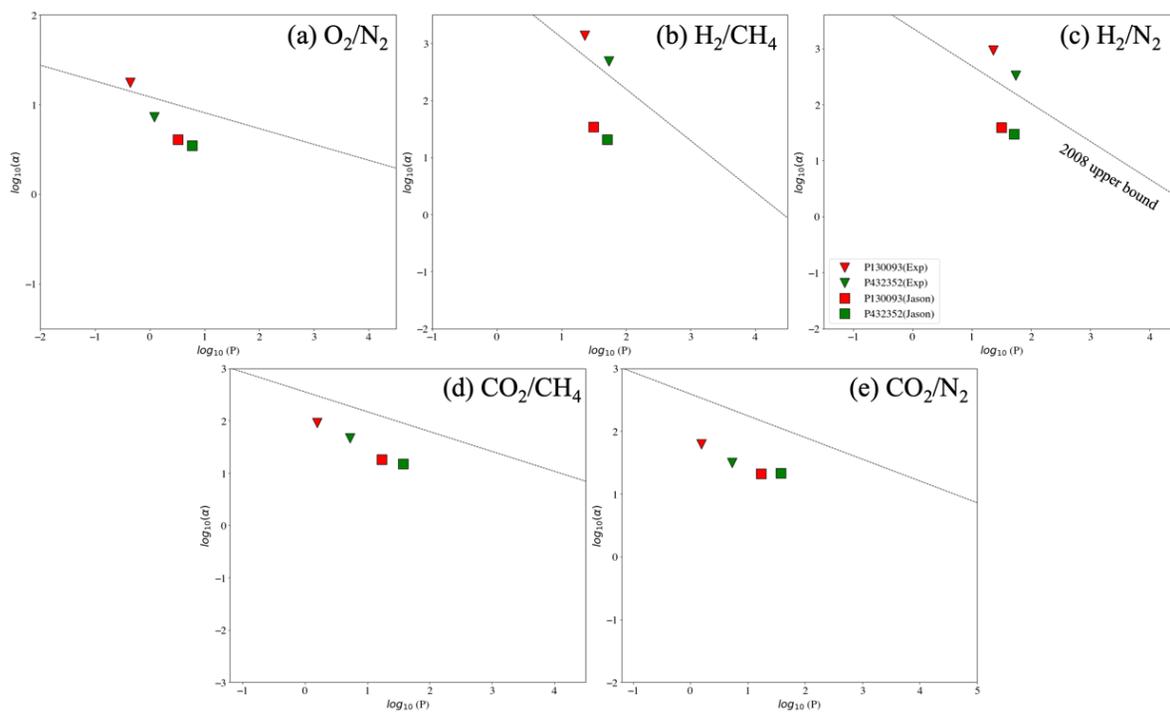

**Figure S10. Robeson plot visualization for five different separation tasks using experiment values and predictions from Jason Yang's models.** Accuracy=5/10.

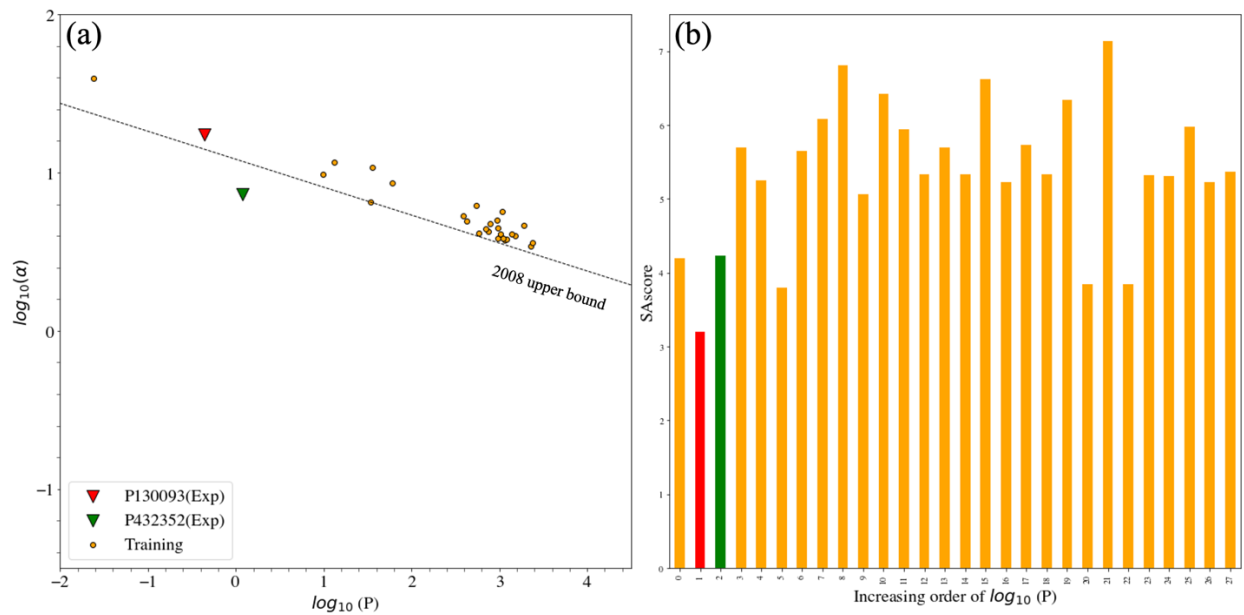

**Figure S11. Polymers in the training dataset with above-the-bound O$_2$/N$_2$ separation performance vs. the two synthesized polymers.** **(a)** Robeson plot visualization for O$_2$/N$_2$ separation; **(b)** SAscore of polymers along the increasing order of log$_{10}$P (left to right). Color orange: training data; red: P130093; green: P432352.

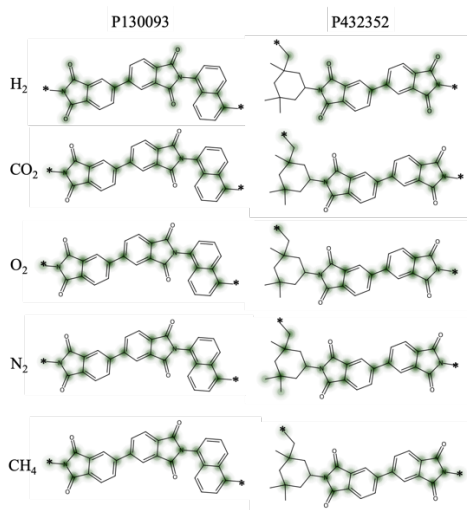

**Figure S12. Molecular graph rationales visualization for P130093 (left column) and P432352 (right column) on five permeability prediction tasks**, including $H_2$, $O_2$, $N_2$, $CO_2$, and $CH_4$ (top to bottom). We visualize the rationale subgraphs using the averaged rationale scores (from 10 independent runs) among the top 50%. The rationale subgraphs are highlighted in green. The darker the color, the higher the rationale score.

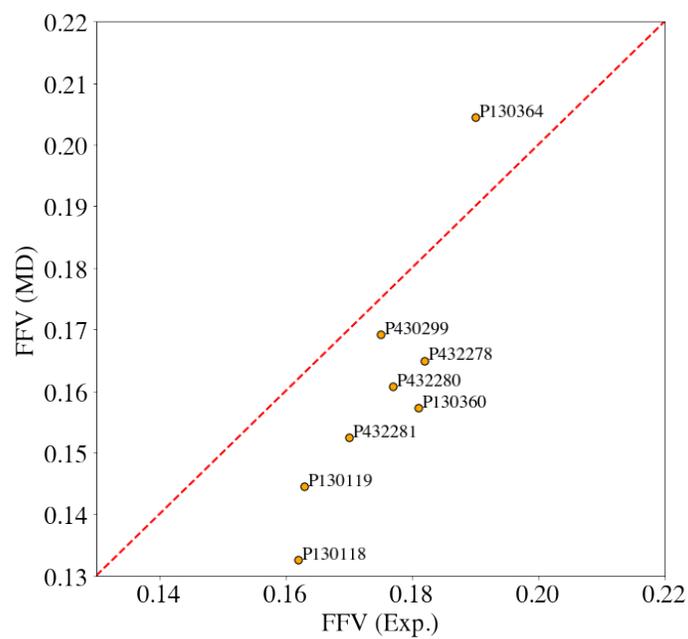

**Figure S13. Benchmark of MD simulated FFV values of 8 polymers from literature** (*9*). Polymers are annotated by their PIDs from PoLyInfo. The MD simulated FFV well captures the trend of experimentally measured FFV.

**Table S2. Comparison of FFV reported in the literature (FFV-Exp)** (*9*) **and FFV calculated by MD (FFV-MD) in this work.** Polymers are annotated by their PIDs from PoLyInfo. The FFV-MD is averaged across the five different snapshots as described in the Methods section. The mean and standard deviation are reported for FFV-MD.

| PID | FFV-Exp | FFV-MD |
| --- | --- | --- |
| *P432283* | 0.19 | 0.141±0.041 |
| *P130364* | 0.19 | 0.204±0.009 |
| *P430299* | 0.175 | 0.169±0.028 |
| *P432278* | 0.182 | 0.165±0.021 |
| *P432281* | 0.17 | 0.153±0.017 |
| *P432280* | 0.177 | 0.161±0.012 |
| *P130360* | 0.181 | 0.157±0.026 |
| *P130118* | 0.162 | 0.133±0.022 |
| *P130119* | 0.163 | 0.145±0.042 |

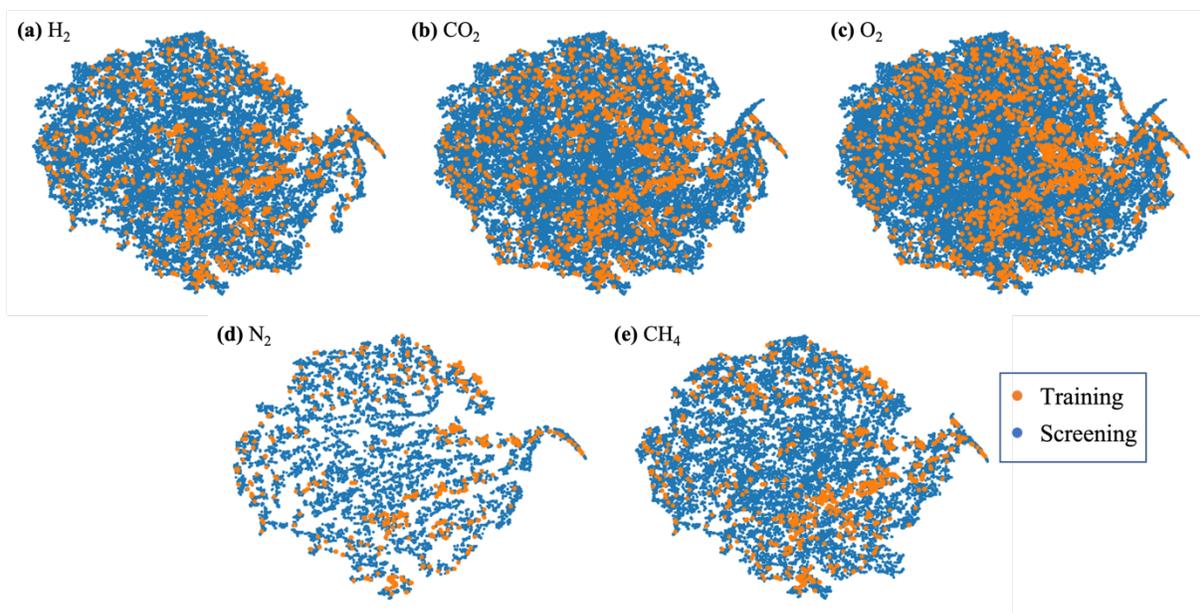

**Figure S14. The t-SNE (*10*) graph embedding visualization for training and screening datasets on five permeability prediction tasks**, including (a) H$_2$, (b) CO$_2$, (c) O$_2$, (d) N$_2$, and (e) CH$_4$. The screening dataset (PoLyInfo) generally spans the feature space of the training data on five different tasks.

**Table S3. Gas permeation test of polymer P130093.** $\left(\frac{dp}{dt}\right)_{ss}$ and $\left(\frac{dp}{dt}\right)_{leak}$ are the steady-state pressure increment in downstream, and the leak rate of the system (cmHg/s), respectively. $l$ is the thickness of the polymer film and $A$ is the effective film area. Measurement errors are shown in parentheses. $H_2$ is not included in the diffusivity calculation because it has a very short lag time and cannot be detected by our existing techniques. $\frac{dp}{dt}$ has a unit of torr/s.

|  |  | $H_2$ | $CH_4$ | $N_2$ | $O_2$ | $CO_2$ |
|---|---|---|---|---|---|---|
| $\left(\frac{dp}{dt}\right)_{leak}$ | | 2.97E-06 | 1.21E-06 | 1.47E-06 | 1.75E-06 | 3.74E-07 |
| $\left(\frac{dp}{dt}\right)_{ss}$ | 30 psig | 1.26E-05 | 1.30E-06 | 1.87E-06 | 9.12E-06 | 3.90E-06 |
|  | 80 psig | 2.43E-05 | 1.74E-06 | 2.19E-06 | 1.74E-05 | 7.72E-06 |
|  | 130 psig | 3.66E-05 | 2.15E-06 | 2.85E-06 | 2.55E-05 | 1.09E-05 |
| Lag-time (s) | | \ | 2667.42 (440.03) | 2508.69 (298.42) | 1281.12 (60.83) | 2385.24 (292.07) |
| $l$ (cm) | | 0.00663 (0.00008) | | | | |
| $A$ (cm²) | | 0.2391 (0.0006) | | | | |

**Table S4. Gas permeation test of polymer P432352.** $\left(\frac{dp}{dt}\right)_{ss}$ and $\left(\frac{dp}{dt}\right)_{leak}$ are the steady-state pressure increment in downstream, and the leak rate of the system (cmHg/s), respectively. $l$ is the thickness of the polymer film and $A$ is the effective film area. Measurement errors are shown in parentheses. $H_2$ is not included in the diffusivity calculation because it has a very short lag time and cannot be detected by our existing techniques. The $\left(\frac{dp}{dt}\right)_{ss}$ of $CH_4$ and $N_2$ at *50 psig* are skipped because of the very slow permeation. $\frac{dp}{dt}$ has a unit of torr/s.

|  |  | $H_2$ | $CH_4$ | $N_2$ | $O_2$ | $CO_2$ |
|---|---|---|---|---|---|---|
| $\left(\frac{dp}{dt}\right)_{leak}$ |  | 1.20E-06 | 2.13E-06 | 1.89E-06 | 1.72E-06 | 4.19E-07 |
| $\left(\frac{dp}{dt}\right)_{ss}$ | *30 psig* | 3.13E-05 | 4.55E-06 | 5.34E-06 | 2.73E-05 | 1.48E-05 |
|  | *50 psig* | 4.43E-05 | \ | \ | 3.85E-05 | 2.10E-05 |
|  | *80 psig* | 6.52E-05 | 7.21E-06 | 9.33E-06 | 5.56E-05 | 2.99E-05 |
| *Lag-time (s)* |  | \ | 2320.96 (123.01) | 606.04 (110.46) | 114.84 (20.21) | 490.12 (36.73) |
| *l (cm)* |  | 0.00408 (0.000056) | | | | |
| *Area (cm²)* |  | 0.1861 (0.00065) | | | | |

# SI REFERENCES